\documentclass[manuscript,screen]{acmart}

\usepackage{graphicx}
\usepackage{multirow}
\usepackage{subfigure}
\usepackage{amssymb}
\usepackage{algorithm}
\usepackage{verbatim}
\usepackage{bm}     
\usepackage{enumitem}       
\usepackage[capitalize]{cleveref}
\usepackage{algpseudocode}
\usepackage{makecell}

\AtBeginDocument{%
  \providecommand\BibTeX{{%
    \normalfont B\kern-0.5em{\scshape i\kern-0.25em b}\kern-0.8em\TeX}}}

\setcopyright{acmcopyright}
\copyrightyear{2018}
\acmYear{2018}
\acmDOI{XXXXXXX.XXXXXXX}

\acmConference[Conference acronym 'XX]{Make sure to enter the correct
  conference title from your rights confirmation emai}{June 03--05,
  2018}{Woodstock, NY}
\acmPrice{15.00}
\acmISBN{978-1-4503-XXXX-X/18/06}

\begin{document}

\title{Negative Sampling in Recommendation: A Survey and Future Directions}


%
\author{Haokai Ma}
\email{haokai.ma1997@gmail.com}
\affiliation{
  \institution{Shandong University}
  \country{China}
}

\author{Ruobing Xie}
\email{ruobingxie@tencent.com}
\affiliation{%
  \institution{Tencent}
  \country{China}
}

\author{Lei Meng$^{*}$}
\email{lmeng@sdu.edu.cn}
\affiliation{%
  \institution{Shandong Research Institute of Industrial Technology; Shandong University}
  \country{China}
}

\author{Fuli Feng}
\email{fulifeng93@gmail.com}
\affiliation{%
  \institution{University of Science and Technology of China}
  \country{China}
}

\author{Xiaoyu Du}
\email{duxy.me@gmail.com}
\affiliation{%
  \institution{Nanjing University of Science and Technology}
  \country{China}
}

\author{Xingwu Sun}
\email{sunxingwu01@gmail.com}
\affiliation{%
  \institution{Tencent}
  \country{China}
}

\author{Zhanhui Kang}
\email{kegokang@tencent.com}
\affiliation{%
  \institution{Tencent}
  \country{China}
}

\author{Xiangxu Meng}
\email{mxx@sdu.edu.cn}
\affiliation{%
  \institution{Shandong University}
  \country{China}
}
\thanks{$^{*}$ indicates corresponding author.}

\renewcommand{\shortauthors}{Haokai Ma, et al.}

\begin{abstract}
Recommender system (RS) aims to capture personalized preferences from massive user behaviors, making them pivotal in the era of information explosion. However, the presence of ``information cocoons'', interaction sparsity, cold-start problem and feedback loops inherent in RS make users interact with a limited number of items. Conventional recommendation algorithms typically focus on the positive historical behaviors, while neglecting the essential role of negative feedback in user preference understanding. As a promising but easy-to-ignored area, negative sampling is proficients in revealing the genuine negative aspect inherent in user behaviors, emerging as an inescapable procedure in RS. In this survey, we first discuss existing user feedback, the critical role of negative sampling and the optimization objectives in RS and thoroughly analyze challenges that consistently impede its progress. Then, we conduct an extensive literature review on the existing negative sampling strategies in RS and classify them into five categories with their discrepant techniques. Finally, we detail the insights of the tailored negative sampling strategies in diverse RS scenarios and outline an overview of the prospective research directions toward which the community may engage and benefit. 
\end{abstract}

\ccsdesc[500]{Information systems~Recommender systems}

\keywords{Recommender System, Negative Sample, Information Retrieval, Survey}


\maketitle

\section{Introduction}
\label{sec.intro}
Recommender systems (RS) have emerged as an effective solution to the information overloading issue, capable of capturing user preference from the massive behaviors and provide appropriate items to each user.
Presently, the extensive deployment of recommendation algorithms underscores their transformative power, shaping the way people engage and navigate the choices with digital content. These encompass electronic commerce platforms (Taobao\footnote{https://www.taobao.com/}, and Amazon\footnote{https://www.amazon.com/}), social network (WeChat\footnote{https://weixin.qq.com/}, and Facebook\footnote{https://www.facebook.com}), lifestyle applications (Meituan\footnote{https://www.meituan.com/} and Google Maps\footnote{https://maps.google.com}) and so on. These significant achievements across multiple platforms and applications underscore the continued significance of RS.






\begin{figure*}[!t]
\centering
\includegraphics[width=0.98\textwidth]{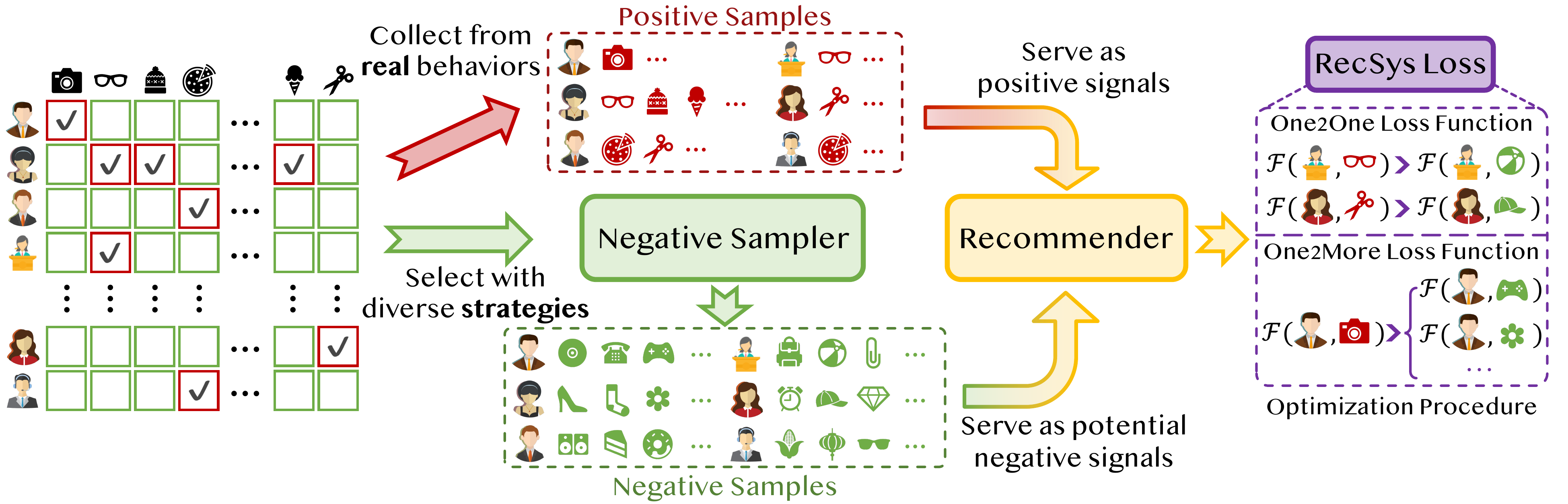}
\vspace{-0.3cm}
\caption{Illustration of the training stage of a personalized recommender system.}
\vspace{-0.4cm}
\label{fig:recommendation_toy_example}
\end{figure*}

Practical RS frequently consists of more than millions of users and items, each characterized by hundreds of auxiliary attributes. Despite this, the existence of personalized preference, ``Information Cocoons'' and the unique feedback loops intrinsic to the RS lead to a scenario in which users only interact with a limited number of warm items, rendering the majority of items of the corpus remain consistently unfamiliar to them \cite{li2022exploratory, lou2019does, chen2021autodebias, meng2020heterogeneous} (as shown in the left part of Fig.\ref{fig:recommendation_toy_example}). The core of RS lies in its ability to infer unobserved preferences from the available interaction data, with a prevailing understanding that their efficacy excels as the richness of interaction data amplifies. However, the interaction data is observational rather than experimental. Thus, the ubiquity of noise within the incomplete users' historical interactions prevalent in real-world RS may lead to the sub-optimal decision-making in recommendation \cite{chen2023bias,gao2023alleviating}.


In alignment with other supervised tasks, providing appropriate supervised signals (both positive and negative feedback) is indispensable in the training phase of recommendation algorithms. The absence of either of these feedback types can inevitably give rise to bias in model training \cite{RealHNS,C16,C24}. However, explicit feedback remains conspicuously absent within most recommendation datasets. Even when encompassed within datasets of such information, certain studies \cite{C32,C23,C24} have unearthed that these negative feedback indeed mirror user preferences to some extent. For instance, the behavior of assigning a low rating to a particular movie may paradoxically signify that the user has already viewed the movie, as this selection from an extensive movie pool subtly reveals the user's underlying predilections \cite{ding2023interpretable,huang2023negative}. Consequently, the integral exploration of negative sampling strategies that are proficient in unveiling genuine negative aspects inherent in user preferences emerges as an inescapable procedure in RS.

As illustrated in Fig.~\ref{fig:recommendation_toy_example}, the training stage of RS typically collect the real behaviors and select the potential negative samples (NS) via diverse sampling strategies simultaneously, and then optimize the recommender algorithm on these given positive and negative signals. The profound impact of distinct negative sampling strategies on the performance of recommendation algorithms in disparate recommendation scenarios underscores the intricacies inherent in this critical domain. Moreover, existing academic endeavors pertaining to negative sampling algorithms are diverse in RS. Varied resolutions are observed in addressing NS across distinct recommendation scenarios, indicating the absence of a universal algorithm capable of concurrently tackling all recommendation tasks \cite{A63,A86,B1,B8,C13,C22,C32,E4,F10}. Finally, the predominant strategies for tackling matching and ranking tasks revolve around the random sampling strategies \cite{A3} and the meticulous integration of genuine negative feedback \cite{E3,E5,F4} for the industrial recommender systems. Currently, there lacks a comprehensive survey that systematically categorizes and presents the existing studies. Such a survey can serve as a handbook to offer a unified exploration of the varied landscape surrounding negative sampling algorithms in RS, thereby addressing the aforementioned issues.



\begin{figure*}[!t]
\centering
\includegraphics[width=0.8\textwidth]{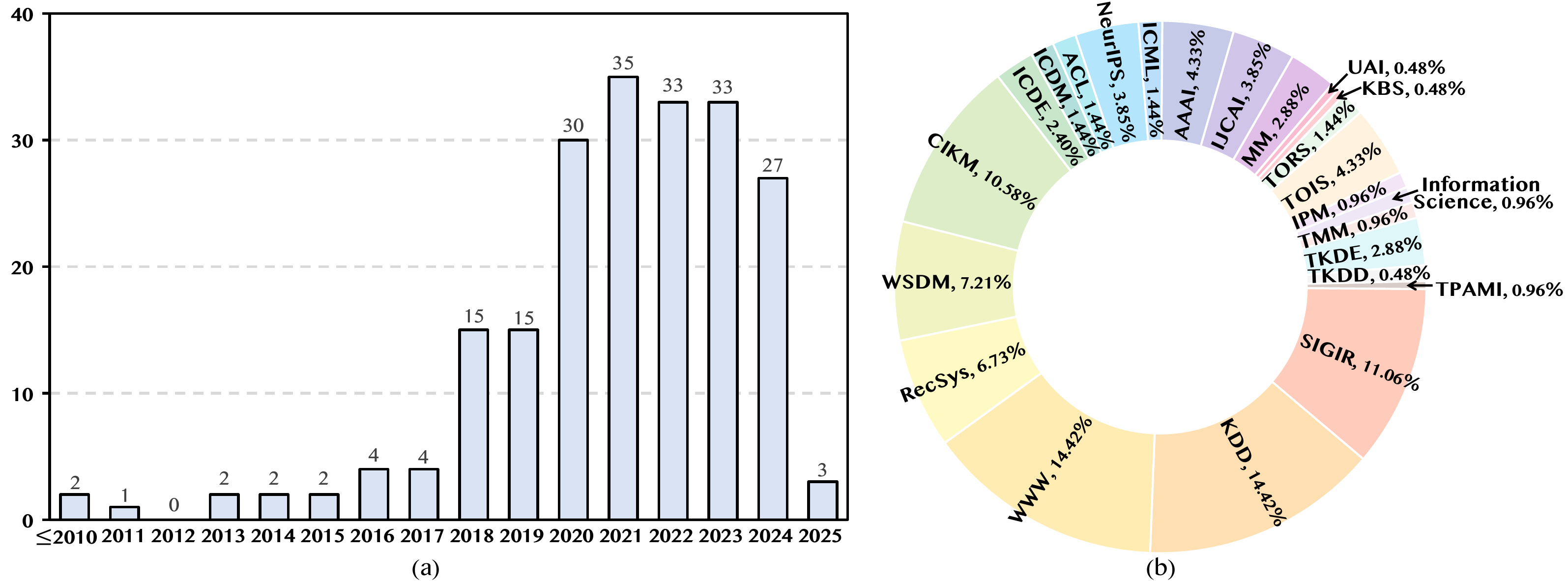}
\vspace{-0.3cm}
\caption{The statistics of publications related to negative sampling in recommendation with the (a) publication year and (b) venue.}
\vspace{-0.4cm}
\label{fig:reference_visualization}
\end{figure*}

Given the significance and popularity of RS, there are several recently published surveys also reviewed this area from different perspectives, including debiased recommendation \cite{chen2023bias}, cross-domain recommendation \cite{zang2022survey}, graph-based recommendation \cite{gao2023survey}, knowledge graph-based recommendation \cite{guo2020survey}, causal inference-based recommendation \cite{gao2022causal}, large language model-based recommendation \cite{Fan2023, Wu2023} and evaluation in recommendation \cite{bauer2023exploring}. As to negative sampling, there are surveys in other domains, including knowledge graph completion~\cite{cai2022temporal}, deep metric learning~\cite{kaya2019deep}, neural semantic retrieval~\cite{guo2022semantic, chen2022deep}, image retrieval~\cite{chen2022deep}, and representation learning on various types of data~\cite{liu2021network, Qian2020NegativeSI, xu2022negative, yang2020heterogeneous, chen2020graph, kamigaito2022comprehensive, wang2021survey, khoshraftar2022survey, wang2017knowledge}. Compared to these research domains, negative sampling techniques employed in RS possess distinctive characteristics, necessitating a comprehensive review and summary. To bridge this gap, we review the negative sampling strategies in RS, classify them into five categories with the discrepant techniques and detail the insights of the tailored negative sampling strategies in diverse recommendation scenarios.

%




This survey encompasses over 230 papers from the top conferences and journals within the field of recommender systems, which focus on the analysis and the design of novel negative sampling techniques. We provide a comprehensive collection of negative sampling strategies in RS from various perspectives. In particular, we search related works from the top-tier conferences on the past 20 years, including SIGIR, WWW, KDD, ACL, AAAI, IJCAI, ICDM, ICDE, RecSys, WSDM, and CIKM, as well as the leading journals such as TOIS, TKDE, TORS, TKDD, KBS, etc, with the keywords ``recommendation'', ``collaborative filtering'', ``ranking'', ``re-ranking'' and ``matching'' in conjunction with ``negative samples'', ``re-weighting'' and ``sampling''. At the same time, to prevent omissions of relevant literature, we utilize Google Scholar, DBLP, and Arxiv to search for related studies. After getting the list of papers, we also check their references to retain the unobserved papers about negative sampling strategies. We provide Figure~\ref{fig:reference_visualization} to illustrate the statistics of our collected publications related to negative sampling in recommendation with their publication year and venue. To facilitate this survey's role as a handbook for negative sampling techniques in RS, we will release a GitHub repository to encompass all reviewed papers along with their corresponding codes at the following link: \url{https://github.com/hulkima/NS4RS}. 

We believe this survey is beneficial for those following academic researchers and industry developers in RS: 1) who are a new to the negative sampling problem and seek a handbook with an overview of related works; 2) who are aspiring to delve into the investigation of negative sampling algorithms but are unsure of where to begin; 3) who are grappling with efficiency or performance issues in their recommendation algorithms and seeking solutions through negative sampling strategies. We summarize the contribution of this survey as follows.
\begin{itemize}[leftmargin=*, topsep=0.2pt,parsep=0pt]
    \item We unify the achievements made in the communities of negative sampling and RS by presenting a comprehensive review of existing related studies, which is the pioneering survey of this promising but easy-to-ignored area. 
    \item We summarize the fundamental challenges of negative sampling in practical RS from three aspects, delineate existing research works into five categories regarding representative methods and characteristics. 
    \item We detail the effectiveness mechanism and meaningful insights of the tailored negative sampling methods in diverse recommendation scenarios and tasks.
    \item We outline further potential research directions for negative sampling in recommendation by intensively analyzing the most advanced technologies in recommendation and the fundamental challenges of negative sampling.
\end{itemize}

\section{Necessity and Challenges of Negative sampling}
\label{sec.nec_and_cha}
Negative sampling emerge as a critical and irreplaceable element in recommendation. Within this specific domain, how to effectively discover users' real negative interests in negative sampling stands as a central research objective \cite{chen2023revisiting}. In this section, our first introduce two types of feedback within recommendation, then we highlight the significance of negative sampling algorithms in recommendation, with particular attention to the crucial role played by hard negative samples, and finally we analyze the existing challenges of negative sampling techniques in recommendation.


\subsection{Introduction of User Feedback in Recommendation}
\label{sec.negative_feedback}

Constrained by the exposure bias and popularity bias inherent in the real-world RS, user interaction reveals a notable unbalancedness, which constitutes a computationally insurmountable problem in recommendation \cite{huang2023negative}. Regarding the available user feedback, both positive and negative feedback can reflect the user's opinions on diverse items in the corpus, which are equally essential for modeling user preferences in RS \cite{A58, wu2022feedrec}. Leveraging such data to its maximum capacity can provide the recommender with a significant information gain. Following the classical studies \cite{E9,C21,E17}, user feedback can be categorized into two different types as explicit feedback and implicit feedback~\footnote{Apart from explicit feedback and implicit feedback, the presence of real negative feedback is a notable factor in practical RS \cite{xie2021deep}. Due to the scarcity of this type of feedback, we discuss their definitions and applications in Sec~.\ref{sec.scenario_mb}.}:

\begin{itemize}[leftmargin=*, topsep=0.2pt,parsep=0pt]
    \item \textbf{Explicit Feedback:} \emph{Explicit Feedback materializes in the form of precise ratings to provides unequivocal insights into user preferences (e.g., from 1.0 to 5.0).}
    \item \textbf{Implicit Feedback:} \emph{Implicit Feedback encompasses data derived from user behaviors without any ratings and reviews (e.g., residence time and whether to choose dislike).}
\end{itemize}

From the above two types of feedback, implicit feedback is fundamentally different from the explicit one, which is captured in a structured manner \cite{saito2020unbiased}. Precisely, explicit feedback delivers direct insights into user preferences, while implicit feedback unravels concealed user interests. Regarding the explicit feedback, existing RS algorithms~\cite{A60,Li,Dai} typically regard the user feedback with lower value (e.g., ratings of [1.0, 2.0, 3.0]) as the negative signals, which may not necessarily reflect the users' real preferences due to the existence of \emph{selection bias} in recommendation~\cite{chen2023bias}. Even for the item explicitly rated as 1.0, they have been chosen by this user from an extensive candidate pool. Therefore, such negative feedback may not be as too "negative" as they reflect this user's preference to some extent. That is, these items appear more "negative" than those the user is interested in, yet somewhat more “positive” when compared to items the user has never interacted with. As noted in \cite{E9}, implicit feedback can be classified into true positives, false positives, true negatives, and false negatives, among which this survey primarily concentrates on the negative aspect. For the negative implicit feedback, traditional RS methods configure all of the uninteracted items as the negative sample pool, which lacks explicitness. Meanwhile, it necessitates the recommender to elucidate users' preferences, which may introduce long-tailed information and exposure bias in RS.

\subsection{Role of Negative Sampling in Recommendation}
\label{sec.nec}
Real-world recommender systems frequently involve more than millions of users and items, rendering the integration of all corpus into the training process prohibitively expensive \cite{C32,C13,F3}. Despite the presence of a substantial number of users and items, the existence of ``Information Cocoons'' and the unique feedback loops within recommendation result in the majority of users interacting with only a limited number of items in the corpus \cite{li2022exploratory,lou2019does,chen2021autodebias}. As a consequence, this gives rise to the well-documented \textbf{issue of data sparsity} in recommendations \cite{PDRec, RealHNS, A53,D6}. The \textbf{dynamic nature} of user interests also necessitates a continuous process of adaptation and updating for recommenders \cite{A3,C21,E2}. Furthermore, the perpetual introduction of new users and items (both devoid of an adequate reservoir of historical interaction data) engenders the notorious \textbf{cold start problem}~\cite{F2,A66}.

Large-scale recommendation applications typically encompass multiple phases to narrow down the candidate items for the recommender, thereby alleviating the aforementioned issues and significantly reducing their computational complexity \cite{huang2023negative,ding2023interpretable}. These algorithms generally require both positive and negative examples to model the users' personalized preferences. Negative sampling is the critical and irreplaceable element in recommendation that could potentially improve the modeling of dynamic user preferences with their sparse interactions. Its crucial secret lies in its ability to select samples from each user's vast collection of unobserved items, specifically tailored to enhance the model's optimization within its current state \cite{C13,C24}. To achieve this, it necessitates the deliberate selection of NS characterized by (1) heightened informativeness, (2) increased discriminative capacity, and (3) enhanced precision. Such samples are typically termed as hard negative samples (HNS), which are derived from their tendency to encompass more comprehensive information compared to the randomly selected samples from the corpus. Incorporating them into the training process serves to balance the positive and negative information within the dataset, thereby ensuring an unbiased optimization of the recommender. Drawing from this foundational definition of HNS, their gradient directions and magnitudes tend to vary from other random NS. Theoretically, more HNS can not only expedite the recommender's convergence but also rectify the optimization direction of the global gradient, thus making it computationally possible to predict the appropriate items for users from the sparse implicit interactions \cite{RealHNS,F6}. In the practical deployment, the inclusion of HNS enables recommenders to elevate their effectiveness and robustness simultaneously, which has been verified in many relevant studies.



\subsection{Optimization Objective in Recommendation}



How to design a effective and robust loss function is the universal but essential research topic in recommendation, which is bidirectionally coupled with the negative sampling techniques in recommendation to jointly establish recommenders' optimization objective and generalization ability. Here, the loss function defines the optimization objectives for positive and negative samples, with different losses exhibiting varying degrees of dependency on and utilization of negative samples. In turn, the quality, quantity, and distribution of negative samples directly influence the gradient direction and the convergence of loss function, potentially undermining its effectiveness. Therefore, a series of studies in recommendation start to investigate effective loss functions to optimize recommenders' convergence towards the global minima. Regarding the ratio between positive and negative samples, these approaches can be broadly categorized into \emph{One2One} (one positive to one negative) and \emph{One2More} (one positive to multiple negatives) schemes.

One of the most representative \emph{One2One} optimization objectives in recommendation is BPR loss \cite{BPR}, which assumes that user prefer the positive item over the non-observed samples. Given user $u$, positive item $i$ and the sampled negative instance $j$, BPR loss $L_{BPR}$ can be formulated as $L_{BPR}=\sum_{(u,i,j)\in \mathcal{D}}-\log\sigma\left(\hat{y}_{u,i}-\hat{y}_{u,j}\right)$, where $\mathcal{D}$ denotes whole training triples, $\sigma(.)$ denotes the Sigmoid function, $\hat{y}_{u,i}$ and $\hat{y}_{u,j}$ represent the score predicted by the recommender. It maximizes the score difference between positive and negative samples, which forces the recommender with relative item preferences rather than absolute scores , thereby closely aligning the goal of top-k recommendation tasks. Compared to the \emph{One2One} optimization objective, \emph{One2More} has attracted increasing attention due to its ability to accelerate recommender optimization and mitigate the impact of potential false negative samples on the global gradient by incorporating several negative samples \cite{SSM, RealHNS}. For instance, BSL \cite{BSL} applies the same Log-Expectation-Exp structure to positive examples and multiple negatives to maintain the robustness of recommenders to the noisy positives. Adap-$\tau$ \cite{Adap} defines the Adaption and Fine-grained principles and utilizes Superloss to adaptively adjust the temperature by monitoring the sample losses for each user. S-DPO \cite{S-DPO} instills ranking knowledge into the LLM via pairing multiple negatives with the positive sample to help LLM-enhanced recommenders distinguish preferred items from negatives. Benefiting from the CL paradigm, AdvInfoNCE \cite{AdvInfoNCE} and PCL \cite{PCL} adaptively assigns hardness to each negative instance in an adversarial manner and collaboratively combine the absolute constraints of BCE/CE loss with CL to make the representations uniformly distributed respectively, thereby mining informative knowledge from the hard negatives. The discussion of optimization objectives in recommendation further underscores the critical role of negative sampling. We believe that the negative sampling strategies summarized in this survey can be seamlessly integrated with existing optimization objectives and will benefit from future advances in this technique to drive progress in recommendation.

\subsection{Challenges of Negative Sampling in Recommendation}
\label{sec.cha}
Real-world recommendation applications still face fundamental challenges with the false negative problem, the trade-off among accuracy, efficiency, and stability, and the universality across different tasks, goals, and datasets.
\begin{itemize}[leftmargin=*, topsep=0.2pt,parsep=0pt]
    \item How to precisely identify negative feedback with the substantial knowledge in recommender systems?
    \item How to balance the accuracy, efficiency and stability in model training?
    \item How to handle multiple scenarios with diverse objectives and data availability in recommendation?
\end{itemize}

\begin{figure*}[!t]
\centering
\includegraphics[width=0.7\textwidth]{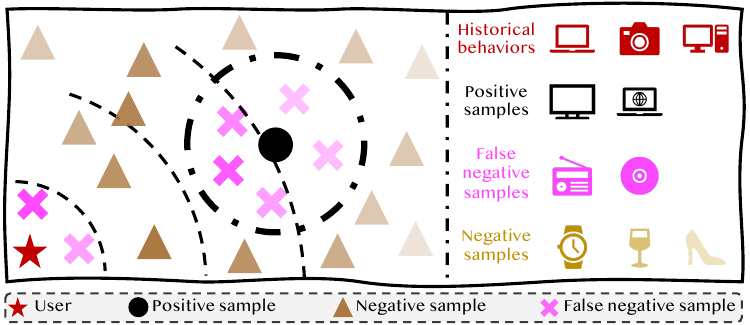}
\vspace{-0.2cm}
\caption{An illustration of diverse types of negative samples in the representation space (Left) and their corresponding toy examples (Right). Here, we utilize the red star, black circle, brown triangle, and pink cross to differentiate the user, positive sample, negative sample, and false negative sample in the representation space, where the brightness of color indicates the "hardness" of each sample.}
\vspace{-0.5cm}
\label{fig:hns_illustration}
\end{figure*}


\subsubsection{False Negative Problem}
\label{sec.cha_fn}
Traditional recommenders randomly select items that users have not interacted with as negative samples \cite{BPR,A46,A58}. In addition to these potentially meaningless random samples, some recommendation algorithms attempt to sample informative yet not excessively challenging negative items from the random candidates \cite{C32,C13,C24}. These samples compel the recommender to delve deeper into modeling the boundary between positive and negative samples in many complex scenarios, thereby improving the recommender's ability to discriminate among the unobserved items. Some recent negative sampling methods~\cite{C13, RealHNS, liu2023ufnrec} have explicitly distinguished these items as the HNS and further given the conception of false negative samples (FNS) in recommendation. We also leverage Fig.~\ref{fig:hns_illustration} to illustrate the distribution and the toy example of each type of samples. For instance, radios and compact discs that share the same category of ``electronics'' may be false negative samples (potential positive samples). Conversely, watches, tall glasses, and high-heeled shoes can serve as the negative samples with the ``hardness'' decreasing, which exhibit less associated with the category of ``electronics''. The corresponding definitions of these samples can be presented as:
\begin{itemize}[leftmargin=*, topsep=0.2pt,parsep=0pt]
    \item \textbf{Negative Samples (NS)}: \emph{As discussed in Sec.~\ref{sec.negative_feedback}, user feedback can be categorized as explicit feedback and implicit feedback. Regarding the former, NS typically denote the items which are rated with low values (e.g., [1, 2, 3]). While for the latter, NS commonly refer the complete set of all items in the corpus that user $u$ has not interacted with. This type of samples typically serves as the fundamental candidate of all existing negative sampling strategies.}
    \item \textbf{Hard Negative Samples (HNS)}: \emph{The samples that possess more information than the majority of \textbf{\emph{NS}}, such as those associated with user $u$'s positive items or items engaged with user $u$'s social connections.}
    \item \textbf{False Negative Samples (FNS)}: \emph{The samples that are erroneously identified as negative samples and subsequently fed into the recommender optimization, which correspond to the real interest of user $u$.} 
\end{itemize}

As highlighted in the discussion of \emph{exposure bias} in~\cite{chen2023bias}, user behaviors are collected from the observed data, which makes them inherently noisy, stochastic, and incomplete. However, the subsequent negative sampling strategies have generally focused on augmenting the "hardness" of HNS. This includes but is not limited to introducing supplementary information \cite{B7,C3,B8,C24}, enhancing the sampling weight of HNS \cite{C32,C13,RealHNS}, and employing positive representation interpolation \cite{C28,C31}.  In this case, certain NS considered overly "hard" by these negative samplers might in fact represent the potential positives~\cite{chen2023bias}. Therefore, incorporating them as the negative samples into the recommender optimization may introduce bias during the back-propagation process, potentially resulting in the poor convergence speed and quickly falling into local optimum and ultimately leading to the False Negative Problem in recommendation. The "hardness" of samples varies significantly across different recommendation algorithms and datasets, resulting in substantial instability in parameter selection \cite{C16,C23,C13,RealHNS} and further aggravating this problem. While some prior works~\cite{MNAR_1,MNAR_2,MNAR_3,MNAR_4} attempt to alleviate this phenomenon by modeling the unbiased representation with the Missing-Not-At-Random feedback, their efforts have largely centered on the theoretical guidance or the refinement of learning objectives, neglecting the more straightforward and effective strategy of negative sampling. To summarize, the false negative problem stands as a prevalent and urgent challenge of this specific research domain to be solved in RS.





\begin{table}[!t]
\caption{Illustration of seven traditional negative sampling strategies with their sampling criteria, time complexities of candidate generation and negative selection, where $\mathcal{I}$ denotes the item set, $\mathcal{R}$ denotes the interaction set, $L$ denotes the number of GNN layers, $C$ denotes the length of candidates, $C_a$ denotes the additional length of memory candidates, $T_c$ denotes the time complexity of computing an instance score, $T_s$ denotes the time complexity of negative selection with a given sampling probability, $T_v$ denotes the time complexity of computing the variance of each instance. Note that $|\mathcal{R}| \gg |\mathcal{I}| \gg C \approx C_a$.}
\vspace{-0.2cm}
\label{table.time_complexity}
\normalsize
\begin{tabular}{c|c|c|c|c|c|c}
\toprule
\textbf{Algorithms} & \textbf{Venue} & \textbf{Year} & \textbf{Scenarios} & \textbf{Sampling criteria} & \textbf{\begin{tabular}[c]{@{}c@{}}Candidate \\ generation \end{tabular}} & \textbf{\begin{tabular}[c]{@{}c@{}}Negative \\ selection \end{tabular}} \\ \midrule
RNS \cite{BPR} & UAI & 2009 & CF & Random sampling & $\mathcal{O(|I|)}$ & $\mathcal{O}(T_s)$ \\ \midrule
DNS \cite{C32}& SIGIR & 2013 & CF & Top-ranked & $\mathcal{O}(C)$ & $\mathcal{O}(CT_c)$ \\ \midrule
PNS \cite{A65} & WSDM & 2014 & CF & Popularity & $\mathcal{O(|R|)}$ & $\mathcal{O}(T_s)$ \\ \midrule
SRNS \cite{C23} & NeuIPS & 2020 & CF & Variance-based & $\mathcal{O}((C+C_a)T_c/E)$ & $\mathcal{O}(C(T_c+T_v))$ \\ \midrule
MixGCF \cite{C28} & KDD & 2021 & GR & Interpolation & $\mathcal{O}(CL)$ & $\mathcal{O}(CLT_c+T_s)$ \\ \midrule
DNS* \cite{C13} & WWW & 2023 & CF & Highest-score scope & $\mathcal{O}(C)$ & $\mathcal{O}(CT_c+T_s)$ \\ \midrule
RealHNS \cite{RealHNS} & Recsys & 2023 & CDR & \begin{tabular}[c]{@{}c@{}}Multi-granularity \end{tabular} & $\mathcal{O}(\mathcal{|I|}T_c/E)$ & $\mathcal{O}(CT_c+T_s)$ \\ \bottomrule
\end{tabular}
\vspace{-0.4cm}
\end{table}

\subsubsection{Trade-off among Accuracy, Efficiency and Stability}
\label{sec.cha_tradeoff}

The "information overload" phenomenon has compelled recommenders to grapple with the issue of striking an appropriate trade-off among accuracy, efficiency, and stability \cite{chen2023bias, gao2022causal}. Negative sampling is no exception to this challenge \cite{C23,A86,B5}. Accuracy pertains to the precision of model's predictions regarding user preferences, which is critical in evaluating the performance of recommenders. Efficiency focuses on how many queries the recommender can process within a unit of time, directly influencing its scalability and user experience. And stability concerns the consistent performance of the recommender across diverse datasets.

Nonetheless, these three aspects are generally interrelated in RS, making it challenging to balance them simultaneously. We furnish the sampling criteria, as well as the time complexities of candidate generation and negative selection for seven classic negative sampling algorithms, arranged in chronological order in Table \ref{table.time_complexity}. This illustrates that with the advancement in the accuracy of negative sampling techniques, there is a synchronous increase in their time complexity, underscoring the delicate trade-off between accuracy, efficiency and stability \cite{BPR,A65,C32,C23,C13,C28,RealHNS}. This is due to the fact that enhancing accuracy may lead to increased computational complexity, resulting in decreased efficiency. Some efficiency-improving optimization techniques may exhibit greater sensitivity to data variations and neglect specific details, potentially impacting the model's accuracy and stability. Simultaneously, the pursuit of stability may increase time and space complexity, conflicting with the balance between accuracy and efficiency. Therefore, achieving the optimal balance among these three factors within negative sampling is a complex and challenging task.

\subsubsection{Universality with Different Tasks, Objectives, and Datasets}
\label{sec.cha_uni}
In essence, the prevailing recommendation algorithms encompass different datasets with diverse distribution, distinct training objectives (e.g., matching and ranking), various recommendation scenarios (e.g., collaborative filtering, multi-modal recommendation, multi-behavior recommendation, sequence recommendation, and cross-domain recommendation), and diverse evaluation methods (random sampled testing, popularity-based sampled testing, and full testing). Although general negative sampling strategies (RNS \cite{BPR}, PNS \cite{A65}, DNS \cite{C32}) are available, some tailored negative sampling techniques designed for specific recommendation tasks can outperform these general methods \cite{A63,A86,B1,B8,C13,C22,C32,E4,F10}. Upon a comprehensive examination of the current negative sampling algorithms, it is evident that none of these approaches can perfectly match all recommendation tasks. In practical implementation, employing different negative sampling strategies for various tasks presents a challenge to the efficiency and stability of the recommender. This challenge underscores the obstacles that large-scale recommender systems must overcome when transitioning to the real-world online deployment.

\section{Literature review of Negative Sampling in Recommendation}
\label{sec.methods}

\begin{figure*}[!t]
\centering
\includegraphics[width=0.95\textwidth]{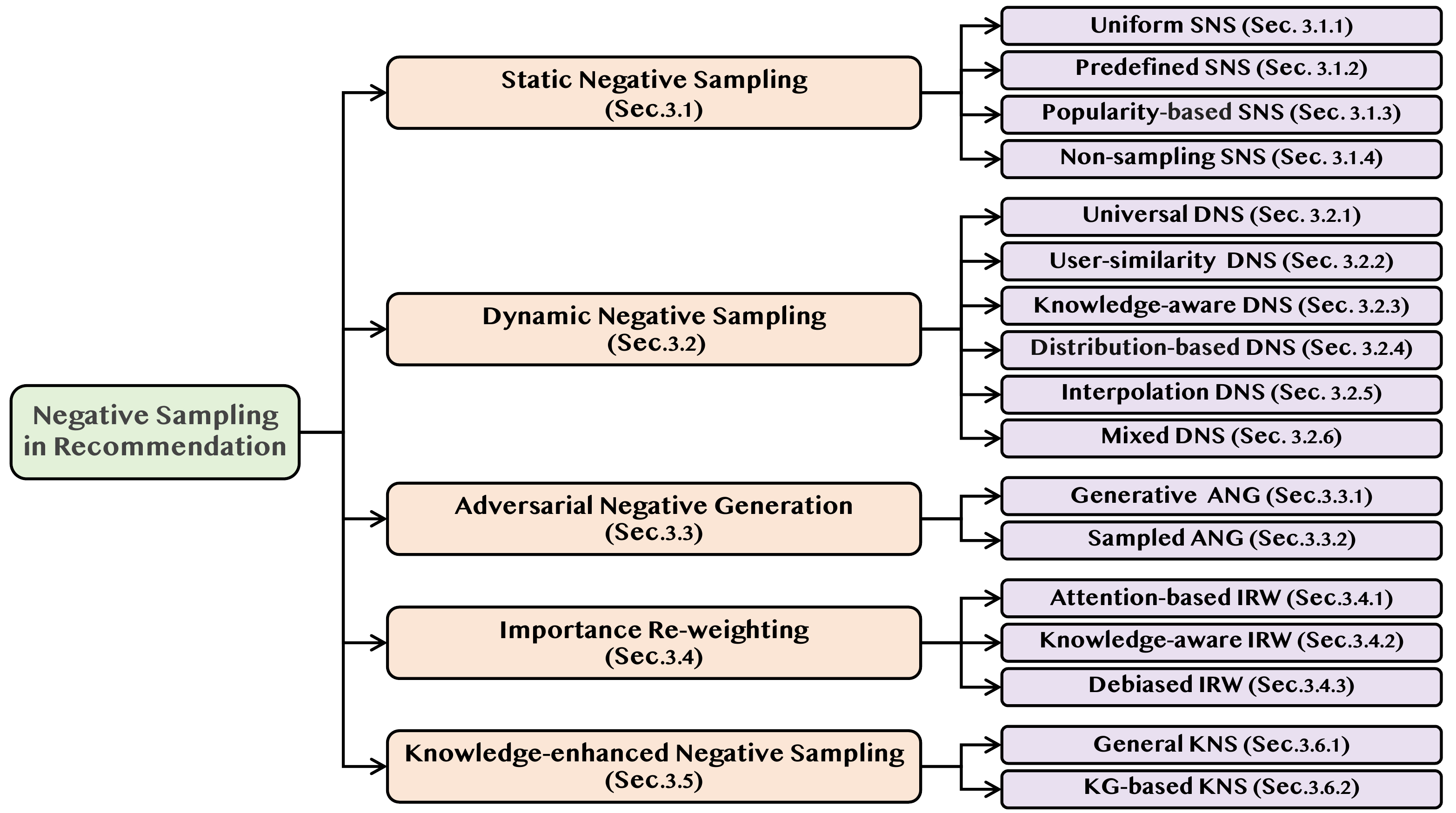}
\vspace{-0.2cm}
\caption{Overall Ontology of Negative Sampling Strategies in Recommendation.}
\vspace{-0.5cm}
\label{fig:tech_classifier}
\end{figure*}

Let $m$ and $n$ be the number of users $\mathcal{U}$ and items $\mathcal{I}$, we can define a user-item interaction matrix $\bm{Y} \!\in\! \mathbb{R}^{m \times n}$ , where $y_{u,i}=1$ denotes the user $u$ has interacted with the item $i$, and $y_{u,i}=0$ denotes $u$ has never interacted with $i$. Negative sampling aims to select the appropriate instances from the unobserved item candidates as the negative samples to support the recommender optimization. We conduct a comprehensive literature review of existing negative sampling strategies in recommendation, classify them into five categories according to their technical trajectories, provide the overall ontology of our taxonomy in Fig.~\ref{fig:tech_classifier} and additionally offer five tables (from Table~\ref{table.sns} to Table~\ref{table.kns}) to illustrate their core contributions, strengths, and drawbacks: 1) Static negative sampling typically samples negative instances with the static probability; 2) Dynamic negative sampling attempts to dynamically select negative instances with the pre-established sampling criterion; 3) Adversarial negative generation leverages the adversarial learning paradigm to sample or generate plausible items as the negative samples; 4) Importance re-weighting aims to identify the importance of each sample and assign diverse weights for them in a data-driven manner; and 5) Knowledge-enhanced negative sampling focuses on sampling negative instances with the mining of the implicit association from the auxiliary knowledge. In the following, we detail the effectiveness mechanisms and meaningful insights of these negative sampling strategies.

\subsection{Static Negative Sampling}
\label{sec.methods_sns}

Traditional approaches in the initial stages of deep learning for recommendation relied on the static negative sampling (SNS) method whereby several items are selected from users' unobserved items. The primary objective of SNS is to offer a diverse range of informative negative samples and facilitate the acquisition of more comprehensive user preference patterns by the recommender. We will introduce several types of SNS strategies (Uniform SNS, Predefined SNS, Popularity-based SNS, and Non-sampling SNS) in the following subsections.

\subsubsection{Uniform SNS}
\label{sec.methods_uniform_sns}
Uniform SNS randomly selects items from the unobserved interactions of the user as the negative sample \cite{A17,dai2024recode}. This strategy is extensively employed across diverse recommendation scenarios (e.g., Collaborative-guided Recommendation \cite{BPR,A83,A39,Graph_RS1}, Sequential Recommendaiton \cite{A31,zou2024knowledge}, Social Recommendation \cite{yang2023generative,yang2024graph,Social_RS1,Social_RS2}, Multi-modal Recommendation \cite{wang2023sequential,wang2024enhanced}, and Cross-domain Recommendation \cite{A53,wang2024making}) and distinct recommendation applications (e.g., Point-of-Interest Recommendation \cite{A7}, News Recommendation \cite{A85}, Song Recommendation \cite{A22} and API Recommendation \cite{A24}). Primarily, its ease-to-deploy mechanism makes it well-suited for large-scale RS, as it obviates the necessity for additional computations. It also introduces a degree of diversity, mitigating the risk of over-fitting and consequently enhancing recommender's generalization and robustness. Moreover, this strategy facilitates the exploration of items that users have not interacted with, thereby aiding the recommender in uncovering novel recommendation opportunities and unexplored areas of user preference.

Notably, TDM \cite{C14} denotes the positive sample as the leaf nodes that have interacted with the user and their ancestor nodes within the hierarchical tree structure, while randomly selecting nodes except positive ones in each level to constitute the set of negative samples. RSS \cite{A74} also tries to propose a recency-based strategy to generate multiple positive samples out of a single user sequence simultaneously, which is orthogonal to negative sampling and can be applied independently. However, given that the primary focus of this survey lies in the examination of negative sampling in recommendation, there has been limited elaboration on the positive-augmented sampling methods. 

\subsubsection{Predefined SNS}
\label{sec.methods_predefined_sns}

Predefined SNS incorporates the pre-defined negative samples from the dataset into the recommenders's training process \cite{A54,A56,A60,zhang2024practical}. For example, DRN \cite{A57} employs real user behaviors (e.g., skipped, clicked, and ordered actions) within the dataset to delineate positive and negative samples. Additionally, a series of studies designate the ratings of $(4, 5)$ in the explicit user interactions as positive samples. Within this context, ELLR \cite{A54} and IF-BPR \cite{A56} segregate ratings of $(1, 2)$ as negative samples, while SCoRe \cite{A59}, ML-MGC \cite{A60} and many LLM-based recommenders \cite{lin2023rella,Xi,Dai,Li} extend this categorization to include ratings of $(1, 2, 3)$ as negative samples.

\subsubsection{Popularity-based SNS}
\label{sec.methods_popularity_sns}

Relying on the assumption that the popularity of items may demonstrate users' global preferences in the specific domain, popularity-based SNS typically selects negative samples based on the popularity of items, that is, the more popular the item is, the more likely it is to be selected as the negative sample \cite{A61, A63, A64, A65, A66, A67, A68, E2, yang2023hyperbolic}. 
This encourages the recommender to incorporate more popular items as negative samples into the training process. For quick updating the recommender parameters with the new incoming data, eALS \cite{E2} assigns the weight of missing data based on the popularity of items by an incremental update strategy. SoftRec \cite{A63} allows some unobserved items to have non-zero supervisory signals by generating soft targets with popularity distribution, thereby incorporating more unobserved knowledge into the training process. Similar to RSS \cite{A74}, SAE-NAD \cite{A64} adopts the general weighting scheme to assign higher confidence to each positive sample based on its frequency. In contrast, it assigns the weights of all negative examples to the same value (e.g., 1.0) for the Point-of-Interest recommendation task. 

Notably, KGPL \cite{A66} and MKR \cite{A67} employ the Popularity-based SNS on the graph structure. For instance, KGPL \cite{A66} directly formulates the sampling confidence of negative samples based on their interaction frequencies across the entire dataset. Furthermore, it assigns reliable pseudo-labels to the interacted items of cold-start users by considering the reachable items under given meta-paths (i.e., the paths rooted at a user's interacted items in KG). Different from the above studies, MKR \cite{A67} follows the negative sampling strategy of the classical work \cite{NS_NLP} in NLP, which reduces the sampling probability for the high-frequency items and augments it for the low-frequency items. This is motivated by the observation that less probable words tend to be more pivotal, while higher probability words such as `the', `a', and `to' bear less weight in NLP tasks. It actually deviates from the core assumptions in recommendation, thus MKR adopts such a sampling strategy primarily to enhance the computational efficiency.

\subsubsection{Non-sampling SNS}
\label{sec.methods_nonsampling_sns}

Non-sampling SNS considers the unobserved instances from the whole training data for recommender learning, thus avoiding negative sampling \cite{A78, A79, No_Sampling, A87, A86}. This strategy argues that negative sampling are highly sensitive to the data distribution and the number of negative samples, making them difficult to achieve the optimal performance in large-scale RS. To converge to a better optimum, EATNN \cite{No_Sampling} first devises an efficient optimization method to learn from the whole training set without negative sampling, and reformulates a square loss function with rigorous mathematical reasoning for accelerating the training process in social-aware recommendation. ENMF \cite{A79} stands as an extended edition of EATNN \cite{No_Sampling} that extends its efficient learning method with the item-based and alternating-based forms, enabling its application for implicit feedback. Due to the behavior characteristic in multi-bahevior recommendation, EHCF \cite{A78} links the prediction of each type of behavior in a transfer manner and applies non-sampling optimization for the recommender by considering all samples in each parameter update. Inspired by the memorization strategies, ENSFM \cite{A87} employs the non-sampling technique to maintain the convergence into a better optimum in a stable way. JNSKR \cite{A86} refine the existing non-sampling strategy into the modeling of the knowledge graph  embeddings, which consists of entity-relation-entity triplets. Furthermore, it devises an efficient optimization approach to learn more precise entity embeddings and item representations within the knowledge graph, resulting in a stable enhancement of recommendation performance.


\begin{table}[!t]
\caption{Illustration of four types of commonly-used Static Negative Sampling  algorithms.}
\vspace{-0.2cm}
\label{table.sns}
\normalsize
\begin{tabular}{m{5.0cm}<{\centering}|c|c|c}
\toprule
\makecell[c]{\textbf{Algorithms}} & \begin{tabular}[c]{@{}c@{}}\textbf{Behavior} \\ \textbf{dependency}\end{tabular}  & \textbf{Advantages} & \textbf{Challenges} \\ \midrule
Uniform SNS \cite{BPR,A74,wang2023sequential,wang2024enhanced,zou2024knowledge} & $\times$ & \begin{tabular}[c]{@{}c@{}}Easy-to-deploy; \\ Time-efficient \end{tabular} & Accuracy \\ \midrule
Predefined SNS \cite{A54,A56,A57,lin2023rella,Xi,Dai,Li} & User-centric rating  & \begin{tabular}[c]{@{}c@{}}Authentic \end{tabular} & Data availability \\ \midrule
Popularity-based SNS \cite{A61,A63,A66,A67,A68,E2} & \begin{tabular}[c]{@{}c@{}}Interaction \\ frequency \end{tabular} & \begin{tabular}[c]{@{}c@{}}Publicly \\ preferred \end{tabular}  & \begin{tabular}[c]{@{}c@{}}Popularity bias; \\ Conformity bias \end{tabular} \\ \midrule
Non-sampling SNS \cite{A78, A79, No_Sampling, A87, A86} & $\times$ & \begin{tabular}[c]{@{}c@{}}High accuracy; \\ Corpus-visible \end{tabular} & Efficiency \\ \bottomrule
\end{tabular}
\vspace{-0.4cm}
\end{table}

\noindent
\emph{Summary:} The in-depth investigation of behavior dependencies, advantages, and challenges of these above four SNS categories are documented in Table \ref{table.sns}. Among them, Uniform SNS emerges as the most widely employed SNS method with its easy-to-deploy and time-efficient character. However, the randomness inherent in sampling introduces variability to its recommendation performance. Predefined SNS leverages authentic user-centric ratings for negative sample selection. Despite its straightforward implementation, it is critically reliant on the accessibility of user behaviors. In the incorporation of the mainstream preferences of the majority of users into the training process, Popularity-based SNS also introduces the inherent popularity bias and conformity bias. Non-sampling SNS incorporates the unobserved items of each user from the whole training data. Maintaining visibility over the entire corpus of the dataset enhances the recommender's performance, yet encountering challenges in computational efficiency.

\subsection{Dynamic Negative Sampling}
\label{sec.methods_dns}


To capture the informative negative samples in a universal manner, a series of Dynamic Negative Sampling (DNS) strategies has sprung up in recent years. DNS refers to selecting the item that is more relevant to the positive sample representations (or user representations) as the negative samples from the dynamically selected item candidates. For example, the pioneering dynamic negative sampling strategy \cite{C32} generally selects the hardest item from the randomly selected item candidates. The subsequent work primarily focuses on modeling the relationship between users and items, or precisely tuning the sampling difficulty of the candidate set. Particularly, DNS have outperformed most of the existing negative sampling methods in different recommendation objectives and scenarios. We categorize existing DNS strategies into six groups: Universal DNS, User-similarity DNS, Knowledge-aware DNS, Distribution-based DNS, Interpolation DNS and Mixed DNS. In the subsequent parts, we will introduce their implementation details and effectiveness mechanisms in detecting informative negative samples and facilitating recommender optimization. 


\subsubsection{Universal DNS}
\label{sec.methods_universal_dns}
Universal DNS selects the top-ranked item as the negative sample from the randomly selected item candidates \cite{C1,C12,C13,C19,PageRank,C27,RealHNS,PDRec,lai2024adaptive}, which is proposed to select more informative negative samples with the current model state and has been widely applied in various recommendation scenarios.

\noindent
\textbf{Top-ranked universal negative selection.} DNS \cite{C32} hypothesizes that any unobserved item should not be ranked higher than any observed positive item and subsequently oversamples items top-ranked by the recommender from the randomly selected candidates. Due to the fact that the more popular an item is, the more times it acts as a positive sample in the conventional pairwise training procedure. DNS is expected to alleviate the popularity bias in RS.
Along this line, a series of studies \cite{C18,C22} enhance DNS from different perspectives. For example, MCRec \cite{C18} pretrains the nodes' embedding with a pure matrix factorization method, computes the pairwise similarities between two consecutive nodes along a path instance, and then keeps top-K path instances with the highest average similarities. To efficiently identify hard negative instances, CA-FME \cite{C20} samples negative instances from triple sides of the positive triplet, and grants larger sampling probability for harder negative samples according to their compatibility scores.

\noindent
\textbf{Dynamic negative selection with high-ranking scope.} PinSage \cite{C19} argues that the highest-scored item is likely to be a potentially positive sample, leading to the false negative problem. Meanwhile, it also demonstrates that top-ranked items are more similar to positive items than randomly selected ones, which forces the recommender to learn to distinguish items at a finer granularity. Specifically, it utilizes the query items' PageRank score to rank the uniformly selected candidates, randomly samples hard negative items from the items ranked between 2000 and 5000. ITSMGCN \cite{C22} proposes an negative sampler to detect the best sampling scheme by carefully tuning the number of item candidates and the ranking position of hard negatives in ranking diagram. DNS$*$ \cite{C13} further reduces the probability of selecting overly hard negative samples by expanding both the number of item candidates and the selection range in DNS. 

\noindent
\textbf{Collaborative distillation of dynamic negative samples.} To address the sparsity of positive feedback and the ambiguity of missing feedback in recommendation, CD-SG \cite{C29} follows the knowledge distillation (KD) paradigm and proposes collaborative distillation (CD). 
Specifically, the student model of CD-SG samples items with the dynamic soft target according to their ranking orders in the teacher model, with the aim to capture both positive and negative correlations and overcome the disadvantage of rank distillation that ignores negative feedback among items. Similarly, DE-RRD \cite{C26} reformulates the task of learning all the precise ranking orders in recommendation to a relaxed ranking matching problem within the framework of KD. It first defines both the detailed ranking orders among the interesting items and the relative ranking orders between the interesting items and the uninteresting items from the teacher’s recommendation list and then defines a relaxed permutation probability to locate all the interesting items higher than all the uninteresting items in student model, while maintaining the detailed ranking orders among the interesting items.

\noindent
\textbf{Dynamic negative selection with representation learning.} With the advancement of representation learning in recommendation, a cohort of researchers have embarked on an investigation of the distinction between easy and hard negative samples from the perspective of representation learning.
To illustrate, PDRec \cite{PDRec} examines the inherent information disparities among observed instances (positive samples) and the latent relevance within the unobserved items (negative samples). It believes that following the emphasis on information-rich samples within the corpus as the positive instances, the deployment of hard negative sampling may concurrently exacerbate the data imbalance and over-fitting risks within recommenders.
To address this issue, PDRec designs a noise-free negative sampling strategy to capture items with low confidences as the \textbf{\emph{Safe Negative Samples (SNS)}} which exhibit no relevance to user preferences in model optimization. 
In contrast, RealHNS \cite{RealHNS} meticulously investigates the definition of \textbf{\emph{Real Hard Negative Samples (RHNS)}} (samples that are neither too easy nor too hard) and \textbf{\emph{False Hard Negative Samples (FHNS)}} (samples that represents users' actual preferences in the future) and then defines the equation: \textbf{\emph{HNS}}=\textbf{\emph{RHNS}}+\textbf{\emph{FHNS}}. 
To effectively leverage these two concepts, RealHNS employs a curriculum learning framework to construct the multi-granular RHNS selector (coarse-grained candidates selector and fine-grained samples selector). This facilitates the dynamic sampling of high-quality RHNS that are suitable for recommender's current state, thereby alleviating the issue of negative transfer.


\subsubsection{User-similarity DNS}
\label{sec.methods_u_sim_dns}

User-similarity DNS identifies similar users based on their historical behaviors and then dynamically selects negative samples according to this similarity association \cite{C1, C3, C4, C33}. 
Some studies emphasize the explicit associations among users to approximate users' conditional preference state, inferring the information dissemination path and collaborative relationships through the connections in social networks or explicit interaction behaviors. SamWalker \cite{C33} proposes an social-based sampling strategy to both speed up the gradient estimation and reduce the time-consuming of the preference inferring from the massive unobserved feedback. To reduce the dependency on other users' preferences, EvalRS \cite{C4} proposes a relatively audacious negative sampling strategy to capture neighboring users with their collaborative representations and regards item from the interaction list of the user's neighbors that the user have not interacted as the negative sample. 
Conversely, other studies focus on implicit relationships, which refer to connections that are not directly observable. Such approaches aim to reveal the potential collaborative behaviors and interest similarities by mining users' behavioral patterns. DEEMS \cite{C1} believes that unobserved positive samples are more likely to exist correlations with observed positives than unobserved negatives. Based on this hypothesis, DEEMS designs a hedging gradient between the user- and item-centered correlations to rerank the unobserved positive samples before the negative ones. To handle the absence of explicit social information, SamWalker++ \cite{C3} constructs the pseudo-social network, that is, similar users are connected with specific item nodes or community nodes. It proposes a subtle sampling strategy based on the random walk along the pseudo-social network to precisely model the triplet confidence and determine which data are used to update parameters.

    
\subsubsection{Knowledge-aware DNS}
\label{sec.methods_know_dns}
Knowledge-aware DNS dynamically captures the informative samples with their attributes or other relevant features and selects the informative items related to the positive sample as negative samples \cite{C2,C24,C6, E10}. This strategy is designed to incorporate task-specific information to intelligently select more relevant and informative negative examples for the specific task. In Location Recommendation, users' negatively-preferred locations are typically mixed with potentially positive ones. However, informativeness differs from sample to sample, so treating them equally in optimization is far from optimal. GeoSAN \cite{C2} designs geography-aware negative sampler to retrieve some nearest locations to the target location as candidates and employ negative sampling based on the uniform distribution or a popularity-based distribution. CITING \cite{C6} proposes a time-aware negative sampling strategy to select the tweets that were posted in close temporal proximity to his/her positive tweets as negative ones in Tweet Recommendation. RPRM \cite{E10} also designs a novel negative sampling strategy to model the agreement of review properties between the users' interacted items and the unseen items. To investigate the Debiased Recommendation, FairNeg \cite{C24} dynamically perceives the group-level unfairness based on their performance disparity during the training process and adjusts each group's corresponding negative sampling probability, intending to equalize all groups' performance.




\subsubsection{Distribution-based DNS}
\label{sec.methods_distri_dns}
Distribution-based DNS analyzes the distribution pattern of positive and negative samples, then dynamically selects informative negative samples that do not overly disturb the training phase \cite{C16,C23,F14}. Some sampling strategies \cite{C23,C16} explore the statistical properties of both \textbf{\emph{RHNS}} and \textbf{\emph{FHNS}} to distinguish the relevant negative samples that are less informative or misleading. SRNS \cite{C23} discovers an empirical conclusion that only a limited number of samples are potentially crucial for the recommender, and false negative samples tend to get stable scores during training and further proposes a variance-based sampling function to distinguish negative samples. GDNS \cite{C16} introduces an expectational gain-tuning sampler to leverage the high-gap expectation of users' preference in the training process to measure the quality of negative samples and dynamically guide the negative sampling procedure. In contrast, DENS \cite{F14} argues that real-world interactions are primarily driven by certain relevant factors associated with the items while not encompassing all factors. Therefore, It delineates the relevant and irrelevant factors within items that are associated with users or not and then devises a factor-aware sampling strategy to sample the appropriate by comparing the relevant factors while ensuring the similarity of irrelevant factors.


\subsubsection{Interpolation DNS}
\label{sec.methods_interpolation_dns}
Interpolation DNS synthesizes the informative negative samples by injecting information from positive samples with the interpolation technology \cite{C28,C31}. These approaches aim to explore the balance between maintaining the informative knowledge while avoiding excessive disturbance to the distribution of the representation space during optimization, allowing recommender to adapt to the changing requirements and maintain the balance between positive and negative samples. MixGCF \cite{C28} interpolates different proportions of positive embeddings into item candidates, optimizing the representation of the selected samples and translating the potential FNS into negative samples to avoid training bias. SDNS \cite{C31} proves that the positive-dominated mixing in MixGCF has the same effect as scaling the sigmoid function in BPR loss function in DNS, and specifically proposes an improvement scheme to define a soft BPR loss function, which not only preserves the integrity of the DNS \cite{C32}, but also yields remarkable improvements in terms of effectiveness and robustness.

\subsubsection{Mixed DNS}
\label{sec.methods_mixed_dns}

Mixed DNS combines multiple DNS strategies to generate more diverse and effective negative sample candidates \cite{C7,C8,C9,C10,C11,C17,C21}. By integrating the uniform negative examples with those above DNS strategies, the existing Mixed DNS strategy provides a flexible framework that allows recommenders to customize the task-specific combination of different DNS strategies and attempt to achieve a more comprehensive and robust negative sampling process. As the pioneering Mixed DNS strategy, NNCF \cite{C7} proposes three novel sampling strategies to significantly improve training efficiency as well as the recommendation performance, named stratified sampling, negative sharing, and stratified sampling with negative sharing. Stratified sampling selects stratum (set of links on the bipartite graph sharing the same source or destination node) based on either item or user. Negative sharing constructs negative pairs by connecting non-linked users and items in the batch without any increasing computational burdens. And stratified sampling with negative sharing denotes the combination of the above two sampling strategies. For Conversational Recommendation, EAR \cite{C8} designs a mixed negative sampling strategy to simultaneously learn the user's general and specific preference with two types of negative samples (the traditional randomly selected samples and the samples satisfying the partially known preference in the conversation). FB-MNS \cite{C11} discovers that models trained with hard negative samples cannot outperform random negative samples. Therefore, It proposes a mixed sampling strategy to blend random and hard negatives with the ratio as $100:1$ in the training process, where the hard negative samples are selected between the rank position 101-500 to achieve the best Recall metric. GRU4Rec$*$ \cite{C9} extends the in-batch negative sampling strategy of GRU4Rec \cite{C10} with the additional shared samples in the mini-batch, where the shared samples are sampled with a popularity-based SNS strategy. MNS \cite{C17} employs the same mixed sampling strategy as GRU4Rec$*$ \cite{C9} that incorporates the randomly selected additional samples as the supplement of the in-batch negative sampling strategy to tackle the selection bias and the inflexibility to adjust sampling distribution. Taking into account user feedback on attributes, FPAN \cite{C21} employs DNS \cite{C32} to select informative negative samples in addition to directly sampling the non-interacted items as negative items, accelerating the training process.


\begin{table}[!t]
\caption{Illustration of six types of commonly-used Dynamic Negative Sampling algorithms.}
\vspace{-0.2cm}
\label{table.dns}
\resizebox{\linewidth}{!}{
\normalsize
\begin{tabular}{m{4.5cm}<{\centering}|c|c|c}
\toprule
\makecell[c]{\textbf{Algorithms}} & \textbf{Sampling criteria} & \textbf{Advantages} & \textbf{Challenges} \\ \midrule
Universal DNS \cite{C1,C12,C13,PageRank,RealHNS} & \begin{tabular}[c]{@{}c@{}}Relying solely on user \\ -item matching scores \end{tabular}  & \begin{tabular}[c]{@{}c@{}}Easy-to-deploy; \\ Universal \end{tabular}  & Accuracy \\ \midrule
User-similarity DNS \cite{C1, C3, C4, C33}	& \begin{tabular}[c]{@{}c@{}}Identifying user \\ similarity association \end{tabular} 	&\begin{tabular}[c]{@{}c@{}}Capturing user \\ behavior similarities \end{tabular} & \begin{tabular}[c]{@{}c@{}}Relies on user \\ associations; Poor \\ performance for new users \end{tabular} \\ \midrule
Knowledge-aware DNS \cite{C2,C24,E10}	& \begin{tabular}[c]{@{}c@{}}Emphasizing samples with \\ same attributes to positives \end{tabular}	& \begin{tabular}[c]{@{}c@{}}Capturing item \\ content correlation \end{tabular} &	\begin{tabular}[c]{@{}c@{}}Additional knowledge \\ dependency \end{tabular} \\ \midrule
Distribution-based DNS \cite{C16,C23,F14}	& \begin{tabular}[c]{@{}c@{}}Analyzing the inherent \\ distribution in dataset \end{tabular} 	& \begin{tabular}[c]{@{}c@{}}Focuses on \\ real negatives \end{tabular} & \begin{tabular}[c]{@{}c@{}}Additional memory \\ dependency \end{tabular} \\ \midrule
Interpolation DNS \cite{C28,C31}	& \begin{tabular}[c]{@{}c@{}}Injecting information \\ from positive samples \end{tabular} 	& \begin{tabular}[c]{@{}c@{}}Balance positive and \\ negative samples \end{tabular} & \begin{tabular}[c]{@{}c@{}}Over-smoothing neglects \\ crucial samples \end{tabular} \\ \midrule
Mixed DNS \cite{C7,C8,C9,C10,C11,C17,C21}	& \begin{tabular}[c]{@{}c@{}}Combining multiple above \\ DNS strategies	 \end{tabular} & Flexible	& \begin{tabular}[c]{@{}c@{}}Complex hyperparameter \\ tuning; Substantial \\ computational cost \end{tabular}\\ \bottomrule
\end{tabular}}
\vspace{-0.4cm}
\end{table}

\noindent
\emph{Summary:} The comprehensive examination of the emphasis, advantages, and challenges of six types of DNS algorithms is detailed in Table \ref{table.dns}. Among these methodologies, Universal DNS stands out as an easy-to-deploy and universal method. However, its reliance on user-item matching scores poses a challenge to accuracy. User-similarity DNS prioritizes identifying user behavior similarities, while facing challenges in user association dependency, particularly for new users. Knowledge-aware DNS emphasizes samples with similar attributes to positives to capture their content correlations, introducing an additional dependency on knowledge. Distribution-based DNS analyzes the inherent distribution in the dataset, focusing on real negatives but adding the dependency on space complexity. Interpolation DNS balances positive and negative samples by injecting information from positives but faces the over-smoothing issue, neglecting crucial samples. Mixed DNS has exhibited significant flexibility by combining multiple strategies, but the complex hyperparameter tuning and substantial computational costs bring new challenges to these strategies.


\subsection{Adversarial Negative Generation}
\label{sec.methods_ans}

Adversarial Negative Generation (ANG) is a sophisticated technique which devise to enhance the robustness and performance of RS. Same as other negative sampling strategies, the core of ANG seeks to address the imbalanced data issue, where positive interactions tend to dominate while real negative instances are relatively sparse. Existing ANG studies can be fundamentally categorized into two distinct paradigms: Generative ANG and Sampled ANG. In the forthcoming sub-sections, we will expound upon these two methods from the existing ANG studies in RS.

\subsubsection{Generative ANG}
\label{sec.methods_generative_ans}

Generative ANG attempts to capture the underlying data distribution and generate the mendacious but plausible samples with a generative model instead of imposing any assumption on existing data. It can maintain the Nash equilibrium between generator and discriminator in generative adversarial network (GAN) \cite{GAN} to \textbf{generate negative samples that are similar to positive samples but have subtle differences}, making them hard to distinguish from positive samples \cite{D1,D2,D3,D4,D5,D6,D7,D8}. Prevailing research in Generative ANS generally employs three methods for generating negative samples: user-specific generation, distribution-associated generation, and content-aware generation. Subsequently, we will elucidate each of them in accordance with relevant works.



\noindent
\textbf{User-specific generation}. 
Generative ANG leverages the adversarial mechanism to generate the targeted negative samples for different users by generators and maximize the objective function between real and generated item representations by discriminators, thereby pushing the recommender to its performance limit. RNS \cite{D4} designs a reinforced negative sampler to generate exposure-alike negative instances and train a better recommender as the predictor through the feature matching technique and the adversarial training paradigm instead of directly choosing from exposure data. Different from RNS, ELECRec \cite{D7} regards recommendation task as a discriminative task rather than a generative task, that is, training the discriminator as the final recommender to distinguish if a sampled item is a ‘real’ target item or not. It defines the generator as an auxiliary model of the next-item prediction task to generative high-quality plausible samples, improving the discrimination ability of the discriminator from the intricate correlations.
    
\noindent
\textbf{Distribution-associated generation}. Generative ANS also adopts the adversarial training paradigm to model the optimal data distribution from the correlation between users and items. For example, DRCGR \cite{D2} generates the relevancy distribution of targeted negative samples for different users with the generator and achieves a better estimation of the relevancy between the generated negative feedback sequence and the relevant positive feedback with the discriminator, thereby automatically learning the optimal recommendation strategies. To learn the true relevance distribution of users and items, PURE \cite{D8} continuously samples on both users and items in the representation space. Specifically, given a positive instance $(u, i)$, the generator of PURE generates a fake item $i'$ that is highly likely to be relevant to $u$ and a fake user $u'$ that is likely to be relevant to $i$ from the random Gaussian noise.

\noindent
\textbf{Content-aware generation}. With the contextual information, generative ANG can generate negative samples that align more closely with the content of positive instances, drawing from the heterogeneous knowledge encapsulated in positive samples and historical user interactions. For new-item recommendation, LARA \cite{D3} designs a generative model with multi-generators to generate the user profile which is not only close to the real profile but also relevant to the given item with the item’s attributes. AFE \cite{D6} defines the generator to take the common features (e.g., user historical behaviors, item features, and contexts) rather than future features for generating “fake” items which may be clicked to confuse the future-aware discriminator, thus preventing the feature inconsistency and overfitting issues.

\subsubsection{Sampled ANG}
\label{sec.methods_sampled_ans}
Sampled ANG tries to fit the underlying distribution as much as possible, and \textbf{sample the most likely items as negative instances from the captured distribution} \cite{D1,D5,D9,D10}. Different from the typical Generative ANG methods, the generator of Sampled ANG works directly select instances from the existing items in the corpus rather than generating item representations. As one of the most famous Sampled ANG works, IRGAN \cite{D9} stands as a pioneering endeavor that incorporates GANs into RS with a game theoretical minimax game which optimizes both generator and discriminator iteratively. The primary objective of its generator lies in strategically selecting negative instances from the candidate pool and intermixing them with positive examples to confound the discriminator. Conversely, its discriminator is designed to distinguish positive samples from the negative instances sampled by the generator, and subsequently provide constructive feedback to refine the generator. To be specific, IRGAN tries to sample negative instances based on the estimated correlation between candidate items and queries vie its generator. Similarly, the generator in GraphGAN \cite{D10} attempts to approximate the underlying connectivity distribution in the graph structure and samples the relevant vertices similar to ground truth's real immediate neighbors to deceive the discriminator. In contrast, its discriminator tries to detect whether the sampled vertex is from ground truth or selected by the generator. NMRN-GAN \cite{D1} also develops a GAN-based noise sampler to adaptively sample the hardest negatives from the randomly selected candidates, which considers both the specific user and the recommender parameters with the policy gradient-based reinforcement learning framework. AFT \cite{D5} proposes a domain-specific masked encoder and a GAN-based two-step feature translation to aggregate the domain-specific preferences from other domains, thereby retrieving the fake clicked samples in different domains for adversarial training.

\begin{table}[!t]
\caption{Illustration of two types of commonly-used Adversal Negative Generation algorithms.}
\vspace{-0.2cm}
\label{table.gan}
\resizebox{\linewidth}{!}{
\normalsize
\begin{tabular}{m{2.7cm}<{\centering}|c|c|c}
\toprule
\makecell[c]{\textbf{Algorithms}} & \begin{tabular}[c]{@{}c@{}}\textbf{Generation} \\ \textbf{strategies}\end{tabular}  & \textbf{Advantages} & \textbf{Challenges} \\ \midrule
Generative ANG \cite{D2,D3,D4,D6,D7,D8} & \begin{tabular}[c]{@{}c@{}}Generating the \\ mendacious but \\ plausible samples\end{tabular} & \begin{tabular}[c]{@{}c@{}}Diverse negative generation; \\ Improved model generalization\end{tabular} & \begin{tabular}[c]{@{}c@{}}Complex training process; \\ Pattern breakdown risk\end{tabular} \\ \midrule
Sampled ANG \cite{D1,D5,D9,D10} & \begin{tabular}[c]{@{}c@{}}Sampling likely \\ negative instances \\ related to positives\end{tabular} & \begin{tabular}[c]{@{}c@{}}Simplify GAN training; \\ Directly leveraging authentic data\end{tabular} & \begin{tabular}[c]{@{}c@{}}Homogenization of negatives; \\ Fail to cover all user preferences\end{tabular} \\ \bottomrule
\end{tabular}}
\vspace{-0.4cm}
\end{table}

\noindent
\emph{Summary:} We systematically outline the generation strategies, advantages, and challenges of each ANG strategy in Table \ref{table.gan}. 
Here, Generative ANG endeavors to leverage GAN to generate the mendacious yet plausible samples, in which the generator and discriminator play a game-theoretical minimax game. The generator attempts to synthesize the plausible negative instances, while the discriminator seeks to distinguish between the genuine and generated samples. Through the iterative adversarial training, the generator becomes increasingly adept at generating mendacious but plausible negative instances, thereby enabling diverse negative generation, improving recommender generalization. However, it faces the risk of pattern breakdown and may lead to the relatively complex training process. 
In contrast, Sampled ANG strategically samples appropriate items from the corpus with the Nash equilibrium state between the generator and discriminator. 
Its advantage lies in guaranteeing the simplified GAN training procedure and the utilization of instances originating exclusively from authentic data. This eliminates any potential training bias stemming from varying levels of GAN's generation complexity. Consequently, it effectively alleviates the imbalance between positive and negative samples in RS, enabling its training on a more balanced dataset. Nevertheless, it encounters challenges such as the homogenization of negatives and the difficulty in covering all potential user preferences.


\subsection{Importance Re-weighting}
\label{sec.methods_ir}
Importance Re-weighting (IRW) is one of the sophisticated statistical techniques employed in Data Mining \cite{E14,E8,E15,IRW_}. Its core revolves around the nuanced adjustment of sample weights to prioritize the significance of certain important negative samples \cite{E18,C15}. 
For the recommendation task where unobserved samples vastly outnumber positive feedback, IRW can be instrumental in rebalancing the real-world dataset by assigning diverse weights to different samples, thus enabling the algorithm to focus its learning efforts on the rare but crucial negative instances \cite{C3,C24}.  Its essential idea is consistent with the memorization effect: recommenders tend to learn accessible user preferences in the early stage of the training process and eventually memorize all complicated interactions. Properly assigning weights to individual samples attunes recommenders to the intricacies of imbalanced implicit datasets, thus recalibrating the learning process for the disparate data distribution. This adaptive weighting not only addresses the challenge of imbalanced data but also fosters the development of more robust and precise recommenders, ultimately bring in superior prediction accuracy. The investigation of IRW strategies classifies them into Attention-based IRW, Knowledge-aware IRW, and Debiased IRW and we will introduce their technical details in the subsequent sub-sections.


\subsubsection{Attention-based IRW}
\label{sec.methods_atten_ir}
Attention-based IRW assigns importance scores and adjusts weights for each item, which could focus on the most relevant items for each user and provide suitable recommendations aligned with the user's personalized interests with the attention mechanism. \cite{E13,E14,dai2024modeling,E18,C15}. 
To efficiently train recommenders with millions of items in the real-world applications, this strategy attempts to sample negative candidates from the uniform distribution and then correct them via importance weighting.
A series of works attempt to \textbf{focus on the safe (easy) negative samples} while overlooking the hard negative samples. To adaptively prune the large-loss interactions, ADT \cite{E9} devises the truncated loss paradigm to discard the large-loss samples (hard ones) with a dynamic threshold and a reweighted loss paradigm to adaptively lower the weights of these samples. UIB \cite{C15} innovatively introduces an auxiliary score for each user to define a personalized decision boundary and individually penalize samples that cross the boundary with a hybrid loss of point-wise and pair-wise paradigm.
Moreover, some studies assume that \textbf{more informative negative samples can contribute more to the gradient}, the magnitude of the gradient becomes much larger so that training can be accelerated. To assist the learning process from the most important historical transitions, DEERS \cite{A58} proposes a prioritized sampling strategy to greedily select the most informative negative sample while ensuring that the probability of being sampled is monotonic in a transition's priority in reinforcement learning. 

Recently, some studies imitate the learning attention in human perception to \textbf{assign different weights to diverse negative samples from simple to complex (whose informativeness is from less to more)}. PRIS \cite{E3} proposes a personalized negative sampler based on rejection sampling to assign larger weights to more informative negative samples, which can contribute more to the gradient of the ranking loss function and accelerate the convergence rate. To address the distribution mismatch, REINFORCE \cite{E12} takes the first-order approximation and ignores the state visitation differences as the importance weights of diverse negative samples with the normalized importance sampling strategy to construct a slightly biased estimator of the policy gradient with lower variance. SGDL \cite{E13} investigates the memorization effect within several representative recommenders' learning process, delineates them into the noise-resistant period and the noise-sensitive period. Then it collects memorized interactions as denoising signals at the noise-resistant period and imposes a weight on each sample loss to enhance the robustness of the noise-sensitive period. GeoSAN \cite{C2} also proposes a weighted BCE loss based on importance sampling to reweight the unvisited locations with the negative probability, enabling more informative locations can be more emphasized even with the uniform negative sampler. 
ContraRec \cite{wang2023sequential} leverages the temperature hyper-parameter $\tau_1$ in the proposed CTC loss to assign diverse attention to negative samples according to their hardness in prediction. 


\subsubsection{Knowledge-aware IRW}
\label{sec.methods_know_ir}

Attention-based IRW studies typically design a reweighting function that maps each sample's score or loss value to its learning weight, thereby enabling the recommender to adaptively learn the information of each negative sample in model optimization \cite{E13,E17}. However, Knowledge-aware IRW argues that users' social contexts and items' heterogeneous information serve as influential determinants in shaping the reciprocal engagements between users and items \cite{yang2023dgrec, E4,E8,C33}. In tandem, these elements covertly impact the confidence of the data. When an item is more popular among this user's social neighbors, the user is more likely to know the item, and his feedback is more attributed to his real preference. Correspondingly, in the situation that user's historical behaviors exhibit a level of coherence in their content, recommender can effortlessly predict the items sharing the similar semantic attributes \cite{E14,C3}. This, in turn, addresses the user's preference for particular content within the contextual paradigm, and this knowledge will be more reliable in deriving user preference.

Within the realm of recommendation, the pre-defined knowledge is typically categorized into user-oriented social knowledge and item-oriented heterogeneous knowledge. Different knowledge-aware IRW methods are commonly utilized to integrate these distinct forms of knowledge into importance re-weighting. 
For the user-oriented social knowledge, SamWalker \cite{C33} simulates item information propagation along the real social network and adaptively specifies individual confidence weights to different interactions based on user's social contexts. To mitigate the reliance on authentic user affiliations, SamWalker++ \cite{C3} introduces the rationale of ``wisdom of the crowds'' to define the pseudo-social network, in which similar users are connected with specific item nodes or community nodes. Under this scenario, SamWalker++ can model data confidence and assign adaptive confidence weight to each interaction without any side information.
Furthermore, for the item-oriented heterogeneous knowledge, PREMERE \cite{E4} proposes a meta-learning-based reweighting strategy to produce the most suitable weight at each optimization state, which can handle the rarity of positive samples and the noisiness of negative samples simultaneously. URL \cite{E14} leverages an auxiliary task to consider user state and context knowledge in reinforcement learning and reweighs each item with a long-term reward to avoid the high variance in policy learning caused by the inverse-propensity weighting. MMT-Net \cite{E8} propose an inverse novelty measure, noted as context-bias, to attenuate the original supervised loss by emulating variance-reduction, thereby inducing a novelty-weighted strategy focused on harder samples as training proceeds. 

\subsubsection{Debiased IRW}
\label{sec.methods_debias_ir}

Debiased IRW identifies and corrects the ubiquitous biases inherent in RS (e.g., popularity bias, exposure bias) and assigns higher weights to items that have been overlooked in the past to deliver more equitable and diverse recommendations \cite{E5,E15,C24}. It operates on the premise that traditional recommenders may inadvertently exacerbate disparities by assigning undue weight to certain categories or user preferences.

By assigning diverse weights to items with various categories, debiased IRW endeavors to counteract any predispositions within traditional recommenders. It can foster more equitable item representation, ensuring that no certain categories disproportionately influence the prediction results. For example, FairNeg \cite{C24} proposes a negative sampling distribution mixup mechanism, which incorporates both importance-aware and fairness-aware sampling distribution to simultaneously strengthen the item representation and item-oriented group fairness. Furthermore, some debiased IRW works also meticulously design and incorporate a suite of fairness metrics and concepts to evaluate the fairness and mitigate the inherent bias. SAR-Net \cite{E5} designs a concept of ``Fairness Coefficient'' to measure each sample's importance and use it to reweigh the prediction of the debiased expert networks, thereby mitigating the data fairness issue caused by manual intervention. IF4URec \cite{E15} introduces the ``Influence Function'' to measure the effect on the estimator when adding a small perturbation on training sample, which could reveal its importance. It is able to utilize ``Influence Function'' to reweigh the training loss of each sample to get less loss in an unbiased validation for debiasing.

\begin{table}[!t]
\caption{Illustration of two types of commonly-used Importance Re-weighting algorithms.}
\vspace{-0.2cm}
\label{table.irw}
\resizebox{\linewidth}{!}{
\normalsize
\begin{tabular}{m{3.8cm}<{\centering}|c|c|c}
\toprule
\makecell[c]{\textbf{Algorithms}} & \textbf{Reweighting functions} & \textbf{Advantages} & \textbf{Challenges} \\ \midrule
Attention-based IRW \cite{E3,E9,A58,E13,E14,E16,E17,E18,C15}  & \begin{tabular}[c]{@{}c@{}}Assigning diverse weights to \\ each item with user attention\end{tabular} & \begin{tabular}[c]{@{}c@{}}Adaptable; \\ Flexibly adjusts weights\end{tabular} & \begin{tabular}[c]{@{}c@{}}Complex; \\ Poor interpretability\end{tabular} \\ \midrule
Knowledge-aware IRW \cite{E4,E8,E14,C3,C33} & \begin{tabular}[c]{@{}c@{}}Identifying item's importance \\ with the external knowledge\end{tabular} & \begin{tabular}[c]{@{}c@{}}Interpretable; \\ Suitable for cold-start items\end{tabular} & \begin{tabular}[c]{@{}c@{}}Additional knowledge\\  dependency\end{tabular}  \\ \midrule
Debiased IRW \cite{E5,E15,C24} & \begin{tabular}[c]{@{}c@{}}Correcting the ubiquitous \\ biases and assigning high \\ weights to overlooked items\end{tabular} & \begin{tabular}[c]{@{}c@{}}Deliver equitable and \\ diverse recommendations\end{tabular} & \begin{tabular}[c]{@{}c@{}}trade-off between \\ fairness and precision\end{tabular} \\ \bottomrule
\end{tabular}}
\vspace{-0.4cm}
\end{table}

\noindent
\emph{Summary:} We have systematically delineated the reweighting functions, advantages and challenges of these above IRW strategies in Table \ref{table.irw}. Here, Attention-based IRW emphasizes assigning diverse importance to each item with user interest attention, making it adaptably and capable of flexibly adjusting weights. However, it faces challenges due to its complexity and relatively poor interpretability. Knowledge-aware IRW focuses on identifying each item's importance with external structured knowledge, offering suitability for cold-start issues but introducing the additional dependency on knowledge. Debiased IRW aims to correct ubiquitous biases, assigning higher weights to overlooked items to provide equitable and diverse recommendations. Nevertheless, it involves the trade-off between fairness and precision.

\subsection{Knowledge-enhanced Negative Sampling}
\label{sec.methods_kns}

Knowledge-enhanced Negative Sampling (KNS) is one of the pivotal negative sampling strategies in recommendation, which leverages the availability of supplementary informative knowledge, such as the auxiliary information (i.e., users' social contexts and items' heterogeneous knowledge) and knowledge graphs, to refine the selection of negative samples. It operates within a broad spectrum of recommendation scenarios, ranging from context-aware recommendation to graph-enhanced recommendation, where knowledge plays a crucial role. The core of Knowledge-enhanced Negative Sampling unfolds in two distinctive dimensions: General KNS and KG-based KNS.


\subsubsection{General KNS}
\label{sec.methods_general_kns}

The rich auxiliary relationships of users and the informative knowledge of items typically imply user preferences and item semantics, which could help overcome the sparsity issue in RS. General KNS selects negative samples based on the existing latent knowledge patterns to better understand the semantics and relations between users and items \cite{B4,B7,B8,E8}. The learning process of the knowledge-based recommendation is grounded on contextual information. For instance, MMT-Net \cite{E8} designs the context-bias $s_{\mathbf{c}}$ to reveal the novel tuples which display uncommon combinations of the interaction and its corresponding context, which can be defined as $s_{\mathbf{c}}=\mathbf{w}_{\mathrm{C}} \cdot \mathbf{c}^{n_{\mathrm{C}}}+b_{\mathrm{C}}$, where $\mathbf{c}^{n_{\mathrm{C}}}$ denotes the final context combinations. To enhance the training effectiveness of the proposed context-bias, MMT-Net develops a negative sampling approximation with two learning objectives to randomly identify the likely item given the user and interaction context and identify the likely context given the user and the item, respectively. Then the context-bias $s_{\mathbf{c}}$ has been used to attenuate the loss as $\mathcal{L}_{(u,\mathbf{c},v)}^{'}\!=\!\mathcal{L}_{(u,\mathbf{c},v)}\!-\! s_{\mathbf{c}}$, thereby inducing a novelty-weighted training curriculum focused on harder samples as training proceeds. SA\_OCCF \cite{B7} leverages sentiment information extracted from user reviews to assign the weight for each negative example rather than the global weighting scheme. Instead of the review information, some works also leverage the informative genre and category knowledge for negative samplings. To list a few, LDA \cite{B4} devises a two-tiered sampling approach that first samples a genre $g$ from the learned genre-distribution $p(g|u)$ for user $u$, and then samples the movie $i$ as the negative sample from the learned distribution $p(i|g)$ regarding the genre $g$. IntentGC \cite{B8} samples negative items in the same leaf category as the corresponding positive item to make the model capable of distinguishing hard items.

\subsubsection{KG-based KNS}
\label{sec.methods_KG_kns}

KG-based KNS leverages the explicit association and high-order correlation among users, items, and other entities in KG to sample negative instances \cite{B1,B2,B5,B6}. KG, which introduces extra relations among items and real-world entities (e.g., item attributes) with the comprehensive structural knowledge, are well known for its potential to enhance recommenders' accuracy and explainability.
To precisely improve the semantic relevance between positive and negative samples and enhance the accessible information of the sampled negative instances, a series of works attempt to introduce the high-order structural coherence and the multi-hop node connectivity into negative sampling in RS \cite{B1,B2,B5}. For example, AnchorKG \cite{B1} design a simple but effective strategy to randomly select the $k$ in $\{1,2,3\}$ and conduct a random walk along a $k$-hop path starting from the positive sample in KG. To minimize computational complexity, PR-HNE \cite{B2} employs unigram distribution to sample items (nodes) in candidates that appear in the KG independently of the presence of other items over each interaction (edge). With the reinforcement learning (RL) paradigm, KGPolicy \cite{B5} samples knowledge-enhanced negative items from the relationships between items and entities in KG, emphasizing KG entities and items linked within two-hop paths and facilitating recommenders' learning process. KGAttack \cite{B6} proposes knowledge-enhanced candidate selection to employ KG for localizing some relevant item candidates with similar attributes. Subsequently, it leverages RL paradigm to obtain the sampling probabilities of each candidate item by calculating the similarity between the current state representations and graph embeddings.

\begin{table}[!t]
\caption{Illustration of two types of commonly-used Knowledge-enhanced Negative Sampling algorithms.}
\vspace{-0.2cm}
\label{table.kns}
\resizebox{\linewidth}{!}{
\normalsize
\begin{tabular}{m{2.4cm}<{\centering}|c|c|c}
\toprule
\makecell[c]{\textbf{Algorithms}} & \textbf{Knowledge source} & \textbf{Advantages} & \textbf{Challenges} \\ \midrule
General KNS \cite{B4,B7,B8,E8} & \begin{tabular}[c]{@{}c@{}}Sampling negative with \\ user preference  patterns \\ and item semantics\end{tabular} & \begin{tabular}[c]{@{}c@{}}Considering user-item knowledge \\ context; Captures broader \\ knowledge in multiple domains\end{tabular} & \begin{tabular}[c]{@{}c@{}}Additional data progressing \\ complexity; Relies on high- \\ quality knowledge data\end{tabular} \\ \midrule
KG-based KNS \cite{B1,B2,B5,B6} & \begin{tabular}[c]{@{}c@{}}Leveraging explicit association \\and the high-order correlation \\among entities in KG\end{tabular} & \begin{tabular}[c]{@{}c@{}}Fully leverages rich \\ knowledge relationships in KG\end{tabular} & \begin{tabular}[c]{@{}c@{}}Depends on KG construction; \\ Complexity of maintaining KG\end{tabular} \\ \bottomrule
\end{tabular}
}
\vspace{-0.4cm}
\end{table}

\noindent
\emph{Summary:} As demonstrated in Table \ref{table.kns}, General KNS exploits all available heterogeneous knowledge to sample the more informative negative samples with richer semantic meaning, thereby outperforming the traditional negative sampling methods. It ensures that the selected negative samples are not only dissimilar to the positive samples but also incorporate meaningful semantic relationships with the positive samples, aligning more closely with users' preferences. However, it relies on high-quality knowledge and additionally introduces data processing complexity. In contrast, KG-based KNS leverages the structured information from KG to enhance the quality of the sampled negatives. By incorporating the user-item interactions to the entities in KG, KG-based KNS can introduce the high-order structural coherence and the multi-hop node connectivity into negative samples, thereby precisely improving the semantic relevance rather than the feature relevance between positive and negative samples. However, it excessively depends on KG construction and faces challenges related to the computational complexity of modeling and maintaining the KG.



\section{Negative Sampling in Multiple Practical Recommendation Scenarios}
\label{sec.scenarios}


The field of recommender systems is replete with a vast spectrum of diverse scenarios, which stems from user demands, historical behaviors, and heterogeneous applications, thereby incorporating the field with a blend of challenges and prospects \cite{gao2022causal, chen2023bias}. Benefits from its superior performance and easy-to-employ character, negative sampling has garnered significant attention from researchers and practitioners in recommendation \cite{C32,C13,D4,E3}. Some researchers have attempted to design a universal negative sampling method that is suitable for all recommendation scenarios \cite{RealHNS}. Nonetheless, the employment, objective, and data availability significantly vary across diverse scenarios, leading to the absence of a negative sampling algorithm that can comprehensively encompass all recommendation scenarios. In this section, we undertake an in-depth exploration of the distinctiveness of various recommendation scenarios, delineating subtle disparities in the negative sampling methodologies under the same scenarios. Ultimately, we conclude with a comprehensive discussion and outlook concerning the paradigm of tailored negative sampling techniques for each specific scenario. Specifically, the subsequent discussions span six pivotal recommendation scenarios in the following subsections: Collaborative-guided Recommendation, Sequential Recommendation, Multi-modal Recommendation, Multi-behavior Recommendation, Cross-domain Recommendation and CL-enhanced Recommendation. These discrete discussions and analyses illustrate the adaptability and customization inherent to negative sampling strategies, a fundamental prerequisite for optimizing recommendation algorithms in diverse recommendation contexts.

\begin{table}[!t]
\caption{Illustration of five representative recommendation algorithms with their corresponding negative sampling strategies, publication venue and year in six recommendation scenarios.}
\vspace{-0.2cm}
\label{table.scenarios}
\footnotesize
\begin{tabular}{c|c|c|c|c}
\toprule
\textbf{Recommendation Tasks} & \textbf{Algorithms} & \textbf{Negative Sampling Strategies} & \textbf{Venue} & \textbf{Year} \\ \midrule
\multirow{5}{*}{\begin{tabular}[c]{@{}c@{}}Collaborative \\ Filtering \end{tabular}} 
& NCF \cite{A83} & Uniform SNS & WWW & 2017 \\ 
& ENMF \cite{A79} & Non-sampling SNS & TOIS & 2020 \\ 
& AHNS \cite{lai2024adaptive} & Universal DNS & AAAI & 2024 \\ 
& TIL \cite{E17} & Attention-based IRW & CIKM & 2022 \\ 
& DNS* \cite{C13} & Universal DNS & WWW & 2023 \\ \midrule
\multirow{5}{*}{\begin{tabular}[c]{@{}c@{}}Graph-based \\ Recommendation \end{tabular}} 
 & DRCGR \cite{D2} & Generative ANG & ICDM & 2019 \\ 
 & IntentGC \cite{B8} & General KNS & KDD & 2019 \\ 
 & JNSKR \cite{A86} & Non-sampling SNS & SIGIR & 2020 \\ 
 & AnchorKG \cite{B1} & KG-based KNS & KDD & 2021 \\ 
& SGCF \cite{A39} & Uniform SNS & WSDM & 2023 \\ \midrule
\multirow{5}{*}{\begin{tabular}[c]{@{}c@{}}Sequential \\ Recommendation \end{tabular}} 
 & DEEMS \cite{C1} & User-similarity DNS & KDD & 2019 \\ 
 & GeoSAN \cite{C2} & Knowledge-aware DNS & KDD & 2020 \\ 
  & ELECRec \cite{D7} & Generative ANG & SIGIR & 2022 \\ 
& PDRec \cite{PDRec} & Universal DNS & AAAI & 2024 \\ 
& ContraRec \cite{wang2023sequential} & Attention-based IRW & TOIS & 2024 \\ \midrule
\multirow{5}{*}{\begin{tabular}[c]{@{}c@{}}Multi-modal \\ Recommendation \end{tabular}}
 & RPRM \cite{E10} & Knowledge-aware DNS & WWW & 2021 \\ 
  & KGPL \cite{A66} & Popularity-based SNS & WSDM & 2021 \\ 
 & POWERec \cite{DONG2024101989} & General KNS & Information Fusion & 2023 \\ 
 & FairNeg \cite{C24} & Debiased IRW & WWW & 2023 \\ 
 & FREEDOM \cite{zhou2023tale} & Uniform SNS & MM & 2023 \\ \midrule
\multirow{5}{*}{\begin{tabular}[c]{@{}c@{}}Multi-behavior \\ Recommendation \end{tabular}} 
& DRN \cite{A57} & Predefined SNS & WWW & 2018 \\ 
 & ATRank \cite{A17} & Uniform SNS & AAAI & 2018 \\ 
& KHGT \cite{xia2021knowledge} & KG-based KNS & AAAI & 2021 \\ 
 & MBHT \cite{A31} & Uniform SNS & KDD & 2022 \\ 
 & MBRec \cite{MBRec} & General KNS & TNNLS & 2022 \\ \midrule
\multirow{5}{*}{\begin{tabular}[c]{@{}c@{}}Cross-domain \\ Recommendation \end{tabular}} 
 & MMT-Net \cite{E8} & Knowledge-aware IRW & SIGIR & 2020 \\ 
 & ML-MGC \cite{A60} & Predefined SNS & IJCNN & 2021 \\ 
 & DDGHM \cite{F17} & Scenario-aware NACL & MM & 2022 \\ 
 & C2DSR \cite{A53} & Uniform SNS & CIKM & 2022 \\ 
  & RealHNS \cite{RealHNS} & Universal DNS & Recsys & 2023 \\ \bottomrule
\end{tabular}
\vspace{-0.4cm}
\end{table}

\subsection{Collaborative-guided Recommendation}
\label{sec.scenario_cf_graph}

Collaborative-guided Recommendation (CR) aims to gather and analyze all the collaborative behaviors for predicting users' potential preferences \cite{BPR}. It typically focuses on projecting users and items into the latent embedding space to reflect their correlation \cite{A83}.
In contrast to other recommendation scenarios, the available data of CR is exclusively restricted in the user-item interactions, which constrains its practical accuracy. Moreover, it is widely recognized that merely considering positive instances is insufficient to capture the intricacies of user preferences \cite{C32,C13}. In light of this, a series of CF methods have investigated how to maximize the utilization of interaction data to sample more informative negative instances \cite{A79,C13,C23,E17}.
\begin{figure*}[!t]
    \centering
    \includegraphics[width=0.5\textwidth]{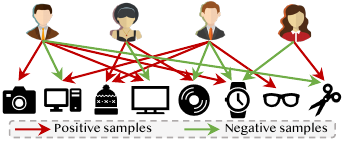}
    \vspace{-0.3cm}
    \caption{An illustration of collaborative-guided recommendation.}
    \vspace{-0.4cm}
    \label{fig:scenario_cf}
\end{figure*}
Constrained by the data availability inherent to CR tasks, prevailing negative sampling methods typically employ SNS approaches to facilitate the process of negative sample selection. Precisely, randomly sampling items from the unobserved corpus with the same probability \cite{A39,A83} are extensively employed in CR. A substantial number of CR approaches also focus on the dynamic selection of negative samples, which are information-rich during the training of CR models. DNS \cite{C32} and DNS$*$ \cite{C13} regard the items with the highest score as the most informative samples and attempt to select them as the negative samples. SRNS \cite{C23}, GDNS \cite{C16}, and DENS \cite{F14} leverage the inherent distribution (e.g., statistical information and disentangled representation) between positive and negative samples to dynamically select the suitable negative instances under the current context in training. Furthermore, a CL-based CR algorithm, EGLN \cite{F12}, attempts to conduct the structural augmentation on the interaction graph to maximize the local-global consistency in collaborative representation learning.



\begin{figure*}[!t]
    \centering
    \includegraphics[width=0.5\textwidth]{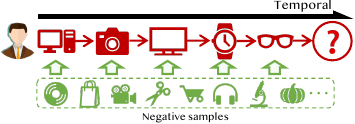}
    \vspace{-0.3cm}
    \caption{An illustration of sequential recommendation.}
    \vspace{-0.4cm}
    \label{fig:scenario_sr}
\end{figure*}

\subsection{Sequential Recommendation}
\label{sec.scenario_sr}
Sequential Recommendation (SR) is a straightforward but effective approach for capturing users' dynamic preference patterns \cite{A52,wang2023sequential,SeeDRec}. Its goal is to predict the next items the user is likely to interacted with by modeling the temporal correlations in his/her behavioral sequence. 
Some studies have proven the effectiveness of hard negative sampling strategies in SR \cite{C1,D7,F19}. Through the selection of negative examples that exhibit a certain similarity to positive samples, these strategies bolster the recommender's capability to finely distinguish users' dynamic preferences.
Most existing SR methods randomly select unobserved items from the corpus as the negative samples \cite{A31,A52,A74}. With the development of contrastive learning (CL), some studies rely on diverse sequence augmentation methods (e.g., crop, mask, reorder, substitute, and insert) to generate negative samples, assisting in modeling users' temporal preferences. Despite the utilization of hard negative sampling strategies, DEEMS \cite{C1} and GeoSAN \cite{C2} opt to treat each interaction with equitable importance rather than explicitly considering the temporal relationships among behaviors
Similarly, ELECRec \cite{D7} leverages discriminator in GAN to distinguish the target sample, overlooking the sequential knowledge within behavioral sequences. Evidently, negative sampling holds particular significance for SR, which has been momentarily disregarded  by  existing researchers and may serve as a promising avenue of exploration for future scholars.


\begin{figure*}[!t]
    \centering
    \includegraphics[width=0.5\textwidth]{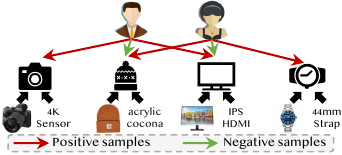}
    \vspace{-0.3cm}
    \caption{An illustration of multi-modal recommendation.}
    \vspace{-0.4cm}
    \label{fig:scenario_mmr}
\end{figure*}

\subsection{Multi-modal Recommendation}
\label{sec.scenario_mm}
Multi-modal Recommendation (MMR) is crucial in real-world recommendation applications with massive modality-specific information, that is, micro-videos, images, audio, and text \cite{wang2023sequential, F25, MCDRec}. 
Since the multi-modal information may reflect users' actual preferences within the fine-grained modality level \cite{C24}, the idealized multi-modal recommender possesses the capacity to uncover the hidden relationships between different modalities and recover the complementary information that can not be captured by CR methods \cite{DONG2024101989}. To achieve satisfactory performance, MMR methods typically require a substantial amount of both positive (present in the observed interactions) and negative items (obscured within the unobserved interactions) to train recommenders in a supervised manner. This motivates researchers to explore how to sample more informative negative instances that better align with users' actual preferences with the multi-modal knowledge.
Similar to other recommendation scenarios, most related studies commonly apply SNS \cite{wang2023sequential,A81,A67} to select negative samples in MMR. Some works also investigate the hidden correlations within the items' multiple modalities to conduct the modality-aware CL pairs. For the traditional negative sampling strategies, FairNeg \cite{C24} achieves debiased recommendation via capturing the group-level unfairness and balancing the sampling probabilities of each semantic group. RPRM \cite{E10} attempts to model the review properties' consistency among different items and consequently sample the negatives with similar review properties of positive samples. POWERec \cite{DONG2024101989} leverages the strength of multi-modal representations learned from the identical item to enhance user interest understanding in the unreliable modality, making all the modality-specific user interests to be fully learned. These algorithms can incorporate the multi-modal information to intelligently refine the selection of the relevant negative instances, which exhibits meaningful semantic relationships with the positive samples and closely aligns with the user's intricate interests.

\begin{figure*}[!t]
    \centering
    \includegraphics[width=0.99\textwidth]{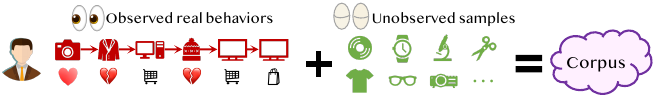}
    \vspace{-0.3cm}
    \caption{An example of the real user behaviors, where \includegraphics[scale=0.1]{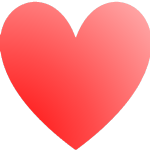} denotes ``like'', \includegraphics[scale=0.1]{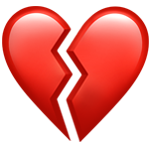} denotes ``dislike'', \includegraphics[scale=0.1]{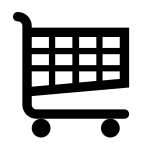} denotes ``add to cart'', \includegraphics[scale=0.1]{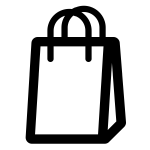} denotes ``purchase'', \includegraphics[scale=0.1]{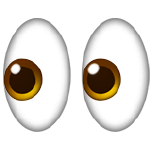} denotes ``observed real behaviors'' and \includegraphics[scale=0.1]{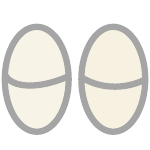} denotes ``unobserved samples''.}
    \vspace{-0.4cm}
    \label{fig:negative_feedback}
\end{figure*}

\begin{figure*}[!t]
    \centering
    \includegraphics[width=0.95\textwidth]{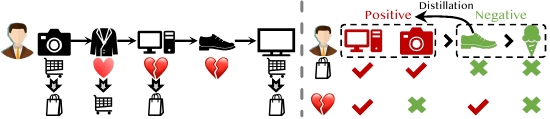}
    \vspace{-0.3cm}
    \caption{An illustration of multi-behavior recommendation.}
    \vspace{-0.4cm}
    \label{fig:scenario_mbr}
\end{figure*}

\subsection{Multi-behavior Recommendation}
\label{sec.scenario_mb}
As illustrated in Fig.~\ref{fig:negative_feedback}, real-world RS typically involves the concept of real negative feedback in addition to implicit and explicit feedback, which contributes to holistic understanding of user preferences \cite{xie2021deep, huang2023negative}. It refers to the authentic user negative behaviors, with its definition intimately linked to various recommendation scenarios. For instance, non-purchases are considered negative samples in the context of purchases, while in the context of clicks, exposures without clicks constitute negative samples. Similarly, non-likes, in comparison to likes, are treated as negative samples. This encapsulates explicit information of users' disapproval, wielding significant influence in recommendation \cite{ding2023interpretable}.

Benefits from the multiple types of interaction between users and items in real-world scenarios, multi-behavior recommendation (MBR) refers to taking full advantage of these heterogeneous user behaviors (e.g., skipped, clicked, and ordered) to alleviate the data sparsity or cold-start issues \cite{xie2021deep, MBRec,F4,zhang2024saqrec}. As illustrated in Fig.~\ref{fig:scenario_mbr}, its core idea is to model the cross-type behavior dependency in RS, which can provide complementary information for user's interest encoding \cite{F4,A17,F29}. In such an application, effectively modeling the multi-type user behaviors can provide auxiliary knowledge and precise user preference to characterize the underlying semantics of user's actual interest representation \cite{A57, F28}. Existing MBR works primarily consider the multi-behavior dependencies with the predefined correlations \cite{A31, MBRec}, while seldom embarking on the selection of diverse negative samples for individual behaviors through negative sampling. To some extent, this limits its ability to capture the intricate cross-type behavior dependencies within complex real-world RS scenarios. For instance, ZEUS \cite{F4} and HMG-CR \cite{F29} independently construct negative CL pairs by forming inter-connections between multiple behaviors and generating behavior-aware hyper meta-graphs. It enables the simultaneous modeling of the diverse interaction patterns and underlying cross-type behavior inter-dependencies, verifying the positive effects of incorporating the cross-type of multi-behavioral context in MBR.





\subsection{Cross-domain Recommendation}
\label{sec.scenario_cdr}

Cross-domain recommendation (CDR) is one of the representative methods to alleviate the data sparsity problem and the cold-start problems in traditional RS with the auxiliary user behaviors collected from other domains \cite{A53,ma2023triple}. It aims to leverage the correlation of users' multi-domain behaviors to accurately transfer the informative knowledge from the source domains and model users' personalized preference in the target domain \cite{A44}. Traditional CDR methods typically conduct the cross-domain knowledge transfer by modeling the feature-level cross-domain correlations with aligned constraint \cite{E8,A43}, adversarial training \cite{A46} and contrastive learning \cite{F17,ma2023triple}. However, these methods merely focus on the randomly-selected negative samples from the target domain, ignoring the cross-domain differences at the sample level. This may disregard users' preference discrepancy between multiple domains and leads to the sub-optimal performance. DDGHM \cite{F17} utilizes random mask operator for negative pair augmentation in CL optimization. Nevertheless, when it comes to recommendation tasks, the approach of random sampling within the target domain is still employed. RealHNS \cite{RealHNS} is the pioneer to systematically discover the false and refine the real from all HNS in CDR. It conducts the multi-grained HNS selectors to find high-quality general and cross-domain HNS with a curriculum learning framework to alleviate cross-domain negative transfer. Notably, there exists substantial uncharted territories for negative sampling methods in CDR, which awaits the investigation by future researchers. This encompasses various areas, including, but not limited to, KNS, ANG, and Debiased IRW.

\begin{figure*}[!t]
    \centering
    \includegraphics[width=0.75\textwidth]{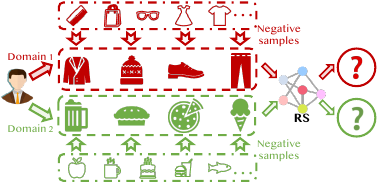}
    \vspace{-0.2cm}
    \caption{An illustration of cross-domain recommendation.}
    \vspace{-0.5cm}
    \label{fig:scenario_cdr}
\end{figure*}

\subsection{CL-enhanced Recommendation}
\label{sec.methods_nacl}
Inspired by the success of CL in computer vision \cite{cl_cv_1}, natural language processing \cite{cl_nlp_1} and graph representation learning \cite{cl_g_1}, some works \cite{F13,F23,F3,tang2024towards} apply CL paradigm in RS. CL aims to maximize the agreement between positive pairs (an item's representation and its corresponding augmented representations) and minimize it between negative pairs (an item's representation and other items' augmented representations) to acquire the transferable knowledge between the original view and the augmented view. Intuitively, CL relies heavily on its negative augmentation (NA) strategy, which is due to the fact that the appropriate augmented view can facilitate the recommender to exploit richer underlying semantic correlation. The most widely used augmentation strategy in CL-based recommendation is the in-batch negative sampling. It treats the augmented representations of the remaining items in the mini-batch where the positive example is located as negative examples, rather than selecting from the corpus, thereby aiding in accelerating its training process without the imposition of additional computational complexity. 

To deeply understand the negative augmentation strategies in CL-based recommendation, we comprehensively dissect its core methodology into four distinct dimensions: pre-defined NA, graph-based NA, sequence-based NA and scenario-aware NA. 
Firstly, pre-defined NA fundamentally aims to refine CL process by augmenting the negative sample pool based on the pre-defined criteria \cite{F11,F18,F28}, which are typically derived from task-specific expertise. For instance, CauseRec \cite{F18} models the counterfactual data distribution by replacing the identified indispensable concepts in user behavioral sequences as negative pairs.
Secondly, graph-based NA refers to generating the negative pairs based on distinct types of graph structures, including interaction graphs, social graphs, and heterogeneous graphs \cite{F6,F9,F12,F16}. It typically perturbs the nodes and edges to provide a nuanced understanding of user-item interactions. SGL \cite{F9}, EGLN \cite{F12}, and RGCL \cite{F13} fully consider the correlations between nodes and edges within the graph structure to propose several graph-based NA strategies, including node (edge) dropout, node (edge) discrimination, random walk, and structure perturbation.
Thirdly, sequence-based NA focuses on augmenting negative samples (e.g., shuffle, mask, crop, etc.) derived from historical behavior sequences \cite{F3,F15,F19,F23}. Its effective mechanism resides in comprehensively capturing the inherent dynamic characteristics of user behavior sequences. With the tailored self-supervised representation augmentation, it effectively handles the sequence structure and builds more robust and accurate recommenders.
Finally, scenario-aware NA involves tailoring the augmentation of negative samples to accommodate the intricacies and demands of various recommendation scenarios. Here, MMR derives multi-view relationships of the user's preferences and item's representation across multiple modalities with CL to avoid the problem of limited diversity or bias \cite{F2,F25}, and MBR constructs negative examples based on the auxiliary behavior data to model the different user behavior patterns through CL \cite{F4,F29}.
Adaptively selecting the appropriate NA strategy based on the scenario requirements and data availability in CL-based recommendation is sensible. Meanwhile, how to construct a universal augmentation method to serve various recommendation scenarios remains an urgent issue.

\section{Conclusion and Future Direction}
\label{sec.future_direction}

Sampling the appropriate negative instances that align with the current model optimization state to model users' unbiased preference is the fundamental yet challenging issue in recommendation. In this survey, we conduct a comprehensive and systematic investigation on negative sampling in recommendation over the past ten years. By holistically analyzing the significance and the challenges of negative sampling research in recommendation, we collect the existing related studies, organize and cluster them into five categories, and outline their diverse sampling schemes to discuss the current research contributions within this area. Subsequently, we systematically detail the diverse insights of the tailored negative sampling approaches in different recommendation scenarios from the perspective of the specific recommendation objective. Finally, we will discuss several important yet not-well-explored research directions in the field of negative sampling to inspire some future studies. We hope this survey is able to offer a comprehensive understanding of this promising but easy-to-ignored area to both newcomers and experts from academia and industry, who are dedicated to negative sampling research in recommendation and to provide some insights for potential future research.


\subsection{Further Explorations on False Negative Sample Issue}
\label{sec.future_direction_fns}


The primary objective of existing negative sampling algorithms in RS is to strategically sample informative yet not excessively challenging negative instances (\textbf{\emph{RHNS}}) from item candidates \cite{C13,RealHNS}. These negative samples are then incorporated to augment the training process of the recommender. From the former studies, we have ascertained the presence of a rather indistinct boundary between \textbf{\emph{SNS}} and \textbf{\emph{HNS}} within the prevailing negative sampling strategies \cite{C12,C16,C23}. This implies that a degree of randomness is inherent in these approaches during negative sampling, rendering it challenging to precisely address the issue of false negatives. Furthermore, users occasionally exhibit preferences for some specific items that diverge significantly from their behavioral patterns, making it difficult to precisely model this occasional preference with user's implicit behaviors \cite{guo2020group, yin2019social}. Following the established nomenclature in RealHNS \cite{RealHNS}, this type of samples is defined as occasional HNS. Regarding the occasional HNS, the utilization of distinct users' multiple behaviors and items' heterogeneous knowledge across different recommendation scenarios may aid in modeling their occasional preference. This could effectively facilitate the utilization or denoising of occasional HNS.

Taking the classical hard negative sampling method (DNS \cite{C32}) as an illustrative example, the size of the candidates effectively determines the ``hard'' level of the sampled negative samples. In fact, the concept of ``hard'' possesses varying definitions across diverse recommenders and even distinct datasets. The indiscriminate employment of DNS to varying recommenders and datasets has been proved infeasible, manifesting the absence of interpretability and stability \cite{D8,chen2021hns,A86}. Consequently, elevating the interpretability, stability, and scalability of existing hard negative sampling strategies while concurrently maintaining the precision remains an open-ended issue and deserves further exploration.

\subsection{Curriculum Learning on Hard Negative Sampling}
\label{sec.future_direction_cl}

Inspired by the empirical evidence in behavior and cognitive science literature, the curriculum learning scheme has been introduced to improve the generalization capacity and convergence rate of neural networks \cite{peterson2004day, bengio2009curriculum}. The core of curriculum learning is to train the model with easier candidates and then gradually increase the hard level of samples until its convergence \cite{wang2021survey}. This process imitates the learning order in human curricula. Its fundamental objective is to enhance the model's learning efficiency by providing more informative training signals to the model as it gradually captures and refines specific knowledge \cite{wang2021survey}. With the assumption that indiscriminately introducing all HNS at the beginning of training may lead to computational wastage, sub-optimal performance, and excessively high gradient magnitudes, some recommendation algorithms attempt to select negative samples with the curriculum learning paradigm \cite{RealHNS, yangunsupervised}. Curriculum learning is able to gradually incorporate increasingly hard training instances in recommender's learning process. It aligns seamlessly with the negative sampling algorithms, which are dedicated to the selection of the most informative samples within the item candidates, those that promise the most significant model enhancement \cite{chen2021curriculum2}. In this situation, negative sampling algorithms and curriculum learning paradigm exhibit similarity to some extent, as they both focus on directing model learning in dynamic and intricate environments \cite{chen2021curriculum1}. Such inherent resemblances present us the opportunity to intricately amalgamate them to guide the training process of recommender more effectively, ultimately culminating in improved performance and enhanced training efficiency.

In the not-so-distant future, it is imperative to explore the integration of negative sampling algorithms with the curriculum learning paradigm. Confronted with the increasingly intricate scenarios and diverse behaviors, the exploration of refining existing negative sampling methods to bolster the seamless collaboration between them and curriculum learning becomes profoundly pivotal. Additionally, the exploration of diverse training strategies holds the potential to achieve heightened model performance and greater efficiency.


\subsection{Causal Inference for Negative Sample Understanding}
\label{sec.future_direction_ci}

Causality is the science of cause and effect in which the effect partly originates from the cause to some extent. As mentioned in \cite{gao2022causal}, causal inference refers to the process of determining and further leveraging the causal relation from the experimental results or observational data. In the recent past, causal inference has been introduced into RS \cite{yang2021top} to extract the causality within the recommenders in modeling intricate user preferences, departing from the traditional correlation-centric paradigm \cite{he2022causpref,he2023addressing}. This paradigm aims to determine how user historical behaviors influence the recommendation results, allowing for a comprehensive comprehension of user preferences.

From the perspective of negative sampling, we posit that the integration of negative sampling with causal inference holds the potential to enable the selection of interpretable and traceable negative samples which can dynamically represent the most suitable instances for recommender. Two kinds of causality widely exist in RS, user-aspect, and interaction-aspect \cite{gao2022causal}. Regarding the user-aspect causality, user-similarity DNS can assist in unearthing commonalities within user behaviors by incorporating causal relationships, thus simulating the driving process behind user decision-making \cite{C1,C3,C33}. In contrast, concerning interaction-aspect causality, universal DNS and distribution-based DNS are proficient at reinforcing users' unadulterated preferences by exploring and establishing causal links between user behaviors and recommended results \cite{C13,C16,C23}. Under these circumstances, negative sample selection is not solely based on the item correlation but also involves the discernment of potential causal relationships among them \cite{chen2019serendipity}. Nevertheless, the integration of negative sampling with causal inference remains an underexplored research avenue in RS, with only a limited number of studies delving into this domain. We believe that it holds great potential to offer the opportunity to provide more accurate, interpretable, and user-centric recommendations.



\subsection{Alleviating Biases in Negative Sampling}
\label{sec.future_direction_ab}

The user behavioral data utilized in the training of recommenders originates from the observed data within the real-world RS rather than experimental data \cite{chen2023bias}. Consequently, this inherent source introduces various biases, which are driven by the diverse exposure mechanisms employed within the real-world large-scale RS. These exposure mechanisms serve as determinants of user behaviors, which, in turn, are subsequently utilized as training data for RS \cite{mansoury2020feedback,wang2023measuring}. Disregarding the aforementioned biases and unquestioningly adhering to the original training strategy would not only introduce additional inherent biases but also lead to the emergence of a "rich get richer" Matthew effect over time \cite{gao2023alleviating}.

To address this issue, some studies attempt to explore debiased learning by adversarial training \cite{adversarial_debias1,adversarial_debias2}, causal modeling \cite{causal_debias1,causal_debias2} and negative sampling \cite{D4,C24}. It is able to improve the fairness of RS by rectifying the impact of biases, whether they stem from data imbalances, user historical behaviors, or various other factors \cite{chen2023bias}. Existing negative sampling algorithms primarily concentrate on mitigating exposure bias, popularity bias, and unfairness, which is achieved by over-sampling popular items \cite{A61,A63,E2} or enhancing the sampler with side information to facilitate model debiasing \cite{C24, B4, B8}. Nevertheless, they have not proposed solutions for the broader issues of selection bias, conformity bias, and other biases in RS. The adaptive negative sampling strategies ensure that the learning process accounts for data imbalances and mitigates any potential biases in RS. Investigating how various negative sampling strategies affect the inherent biases in recommenders would be a genuinely interesting and valuable endeavor.

\subsection{Incorporating Large Language Model into Recommendation}
\label{sec.scenario_llm}

The rapid developments of Large Language Model (LLM) in Natural Language Processing have recently gained significant attention and revolutionized the learning paradigm of diverse research fields \cite{huang2022towards,zhao2023survey,fang2024alphaedit,jiang2025anyedit}. Some studies have indicated that the model size of LLM continues to expand with the immense scale of the corpus size, concomitantly enhancing their capabilities in logical reasoning and generalization \cite{hadi2023large}. Benefits from the potential to bridge the gap between the down-streaming task and the open-world knowledge, LLM have been incorporated into recommendation to serve as the knowledge extractor \cite{wu2021empowering,hou2023learning,hou2022towards} and the recommender \cite{Dai,Gao,Hou,Kang2023,Liu, ma2025large}. Most of the existing LLM-based recommendation algorithms typically employ random sampling \cite{Bao2023,hao2023lkpnr,Mysore2023,Geng,du2024enhancing,wu2024could} or leverage users' explicit scores \cite{lin2023rella,Xi,Dai,Kang2023,Li,lin2023rella} in real-world RS (Ratings within the range of $[4,5]$ are regarded as positive feedback, while defining the rest as negative feedback) for training. The negative samples selected in M6Rec \cite{Cui} are exclusively from real-world systems, that is, Taobao and Alipay. To prevent the excessive training time and the overwhelming influx of negative information that could impact the model's training process, M6Rec further down-samples the negative samples by a factor of ten. PALR \cite{Yang} incorporates the negative samples which are similar to the positive samples (e.g., movies belonging to the same genres and have been co-watched by many users) into model optimization.

From the comprehensive review of existing LLM-based recommendation studies, we notice that LLMRank \cite{Hou} first investigates the ranking performance of LLM on hard negative candidates, which are retrieved by different classical recommendation algorithms. Moreover, InstructRec \cite{Zhang2023} has also delved into LLM's effectiveness and robustness in tasks involving more challenging hard negative samples and larger candidate sets. Nevertheless, these LLM-based recommendation algorithms have not taken into account negative sampling. Furthermore, the prevalent paradigm in these algorithms lies in employing natural language for depicting user interests, context, task formats, and historical behaviors. However, the omission of users' negative preferences in this paradigm introduces the potential bias during the training and inference of LLM, thereby impacting its downstream recommendation performance. A lot of research has demonstrated that the adoption of well-reasoned negative sampling techniques in RS can further bolster the recommenders' performance \cite{C13,D8,E3,F9}. Negative Sampling would serve as a bridge between LLM and recommendation algorithms, facilitating the smooth incorporation of LLM into recommendation tasks. This contributes to being a powerful tool for assisting future researchers in constructing a LLM-based general recommendation framework.

\subsection{Understanding Negative Sampling with Theoretical Tools}
\label{sec.fd_unstt}

To alleviate the inherent sparsity and imbalance in RS, early studies attempt to incorporate suitable negative samples into training from the corpus (auxiliary information or similarity ranking) to provide unbiased negative signals. These works simply attribute their superior performance to the improved convergence of the informative negative samples, without delving further into the profound reasons behind it. Recently, some studies have sought to understand the benefits brought to recommender by negative samples \cite{petrov2023gsasrec} and hard negative samples \cite{C13} through the several theoretical tools. Specifically, gSASRec \cite{petrov2023gsasrec} regards the overconfidence phenomenon of negative sampling as it increases the proportion of positive samples in the training data distribution to force recommenders to overestimate the probabilities of future user-item interactions. It not only results in the recommender's inability to capture subtle variations within high-scored items but also hinders its effectiveness and training convergence. To address this, gSASRec theoretically defines overconfidence through a probabilistic interpretation of sequential recommendation and proposes gBCE loss to mitigate this issue via negative sampling. Inspired by the existing recognition that uniformly sampled negatives may contribute little to the gradients and model convergence \cite{C13, A65, C32, RealHNS}, DNS$^*$ \cite{C13} proves that fast convergence may not be the only advantage of HNS by comparing existing hard negative sampling methods (See Sec.\ref{sec.methods_universal_dns}) with the Non-sampling strategies (See Sec.\ref{sec.methods_nonsampling_sns}). Then DNS$^*$ establish the theoretical foundations and explain the performance gain of HNS by conducting theoretical analyses and simulation studies on One-way Partial AUC. In recent times, these works have demonstrated the superiority of utilizing theoretical tools to understand and guide negative sampling. This proves instrumental in facilitating a deeper exploration of the essence of negative samples in RS and uncovering potential opportunities for performance improvement. It stands out as a promising avenue for future research.


\subsection{Exploring Sampled Metrics in Recommendation Evaluation}
\label{sec.fd_esmre}

Personalized RS aims to leverage users' historical behaviors to capture their interests and provide appropriate items. It is crucial to evaluate the efficacy of these recommenders during their optimization process. In large-scale RS applications, the candidate corpus often comprises millions of items, while it encompasses tens of thousands of items for academic research. Retrieving suitable samples from such an immense corpus poses a considerable challenge. To simplify the computational complexity of offline evaluations, many researchers typically employ uniform random sampling and popularity sampling in evaluating these recommendation algorithms \cite{RealHNS, ma2023triple}. Some recent LLM-based works even verify their effectiveness and robustness in a more challenging setting, that is, evaluating them on hard negative candidates, which are retrieved by different classical recommenders. However, some recent works have proved that these sampled metrics are inconsistent with the full item coverage in algorithm evaluation \cite{krichene2020sampled,A52,pellegrini2022don, bauer2023exploring}.

\cite{bauer2023exploring} is a systematic literature review on the processes surrounding the evaluation of recommendation algorithms, which demonstrates that the predominant evaluation in recommendation algorithms is offline evaluation and that online evaluations are primarily used in combination with offline evaluation. Existing related research has explored the evaluation setting in RS from various perspectives. Precisely, \cite{krichene2020sampled} first explores the implications of sampling during recommender evaluation and discovers the bias in evaluation between random sampling and full item coverage. To alleviate this issue and improve the quality of the sampled metrics, a point-wise correction to the sampled metric is proposed by minimizing different criteria (e.g., bias and mean squared error). \cite{A52} trains four recommenders on five widely known datasets with three evaluation strategies, including full item coverage, popularity-based sampling, and random sampling. It also verifies that the evaluation results obtained from popularity-based sampling do not equal the full item coverage, which is similar to random sampling. \cite{pellegrini2022don} investigates the correlation between optimizing popularity-sampled metrics and estimating point-wise mutual information (PMI) and proposes two PMI fitting techniques to improve popularity-sampled metrics for recommenders. How to precisely evaluate the recommendation algorithms with less computational complexity need to be given more attention for the future research.

\begin{acks}
This work is supported in part by the TaiShan Scholars Program (Grant no. tsqn202211289), the Shandong Province Excellent Young Scientists Fund Program (Overseas) (Grant no. 2022HWYQ-048), the Oversea Innovation Team Project of the "20 Regulations for New Universities" funding program of Jinan (Grant no. 2021GXRC073) and the Young Elite Scientists Sponsorship Program by CAST (2023QNRC001). 


\end{acks}
\bibliographystyle{ACM-Reference-Format}
\bibliography{sample-base}

\end{document}